\documentclass[twocolumn,tighten]{aastex631}
\journalinfo{Accepted version September 7, 2022}
\usepackage{color}

\usepackage{multirow}
\shortauthors{Aruga et al. (2022)}
\shorttitle{Molecular and Atomic Clouds Associated with the Gamma-Ray SNR Puppis~A}

\begin{document}
\title{Molecular and Atomic Clouds Associated with the Gamma-Ray Supernova Remnant Puppis~A}

\author[0000-0001-5069-5988]{M. Aruga}
\affiliation{Department of Physics, Nagoya University, Furo-cho, Chikusa-ku, Nagoya 464-8601, Japan; aruga@a.phys.nagoya-u.ac.jp}

\author[0000-0003-2062-5692]{H. Sano}
\affiliation{Faculty of Engineering, Gifu University, 1-1 Yanagido, Gifu 501-1193, Japan}

\author[0000-0002-8966-9856]{Y. Fukui}
\affiliation{Department of Physics, Nagoya University, Furo-cho, Chikusa-ku, Nagoya 464-8601, Japan; aruga@a.phys.nagoya-u.ac.jp}

\author{E. M. Reynoso}
\affiliation{Instituto de Astronom\'{i}a y F\'{i}sica del Espacio (IAFE, CONICET-UBA), Av. Int. G\"{u}raldes 2620, Pabell\'{o}n IAFE, Ciudad Universitaria, Ciudad Aut\'{o}noma de Buenos Aires, Argentina}

\author[0000-0002-9516-1581]{G. Rowell}
\affiliation{School of Physical Sciences, The University of Adelaide, North Terrace, Adelaide, SA 5005, Australia}

\author[0000-0002-1411-5410]{K. Tachihara}
\affiliation{Department of Physics, Nagoya University, Furo-cho, Chikusa-ku, Nagoya 464-8601, Japan; aruga@a.phys.nagoya-u.ac.jp}

\begin{abstract}
We have carried out a study of the interstellar medium (ISM) toward a shell-like supernova remnant SNR Puppis~A by using the NANTEN CO and ATCA H{\sc i} data. We synthesized a comprehensive picture of the SNR radiation by combining the ISM data with the gamma{-}ray and X-ray distributions. The ISM, both atomic and molecular gas, is dense and highly clumpy, 
%and is distributed along the northeastern edge of the SNR shell. 
{ and is distributed all around the SNR, but mainly in the north-east.}
The CO distribution revealed an enhanced line intensity ratio of CO($J$~=~2--1)/($J$~=~1--0) transitions as well as CO line broadening, which indicate shock heating/acceleration. 
%Further, the velocity distribution of CO and H{\sc i} shows signs of expansion at $\sim$10~km~s$^{-1}$ in the receding part of the shell. 
The results support that Puppis A is located at 1.4~kpc{, in the local arm.}
%.
% with a $\sim$10~$\textrm{pc}$ radius.}
%The results support that Puppis A is located at 2.2 kpc with a $\sim$15~pc radius. 
The ISM interacting with the SNR has {a} large mass of $\sim$10$^{4}$~$M_{\odot}$ which is dominated by H{\sc i}, showing good spatial correspondence with the Fermi-LAT gamma{-}ray image. This favors the hadronic origin of the gamma{-}rays, while additional contribution of the leptonic component is not excluded. The distribution of the X-ray ionization timescales within the shell suggests that the shock front ionized various parts of the ISM at epochs ranging over a few to ten 1000 yr. We therefore suggest that the age of the SNR is around 10$^{4}$~yr as given by the largest ionization timescale.  We estimate the total cosmic ray energy $W_{\rm p}$ to be 10$^{47}$~erg, which is well placed in the cosmic{-}ray escaping phase of an age--$W_{\rm p}$ plot including more than ten SNRs.
\end{abstract}
%X ray X-ray, gamma-ray


%ref, section

\section{Introduction}\label{introduction}
The origin of cosmic rays, mainly consisting of relativistic protons, is one of the longstanding questions in modern astrophysics since their first discovery by {\citet{1912_HESS}}. %Hess(1912).} 
Galactic supernova remnants (SNRs) are thought to be a promising site of cosmic ray acceleration below $\sim$3~PeV (also known as “knee energy”) via the diffusive shock acceleration \citep[e.g.,][]{1978MNRAS.182..147B, 1978ApJ...221L..29B}. %(e.g., Bell 1978; Blandford \& Ostriker 1978). 
The latest observational studies confirmed the acceleration of cosmic ray protons up to $\sim$100~TeV in SNRs \citep[e.g.,][]{2021ApJ...915...84F}. If the supernova origin of cosmic rays is correct, the total energy of accelerated cosmic rays, $W_{\rm p}$, is estimated to be $\sim$10$^{49}$--10$^{50}$~{erg per single} supernova explosion{s}  by considering the frequency of supernova explosion, cosmic-ray energy density, and their confinement time in the Galactic disk \citep[e.g.,][]{2013ASSP...34..221G}. %(e.g., Gabici 2013). 
One of the current challenges is to provide observational support for such theoretical predictions.

Investigating interstellar gas in gamma-ray SNRs holds a key to deriving $W_{\rm p}$ observationally, because the cosmic ray protons emit gamma-rays through p--p collisions with the interstellar protons via $\pi_{0}$--2$\gamma$ process (known as the “hadronic gamma-rays”). Since the hadronic gamma-ray flux is proportional to the target gas density and $W_{\rm p}$, it is essential to accurately estimate the mass of the target interstellar protons. Although the interstellar protons were conventionally assumed to be uniform at density of 1~cm$^{-3}$, such an assumption is not justified as demonstrated by recent radio observations. Several detailed CO/HI radio-line studies revealed {a} highly clumped dense ISM distribution which corresponds overall to the SNR shape, and successfully estimated the mass of the molecular and atomic clouds including the target protons in Galactic/Magellanic gamma-ray SNRs \citep[e.g.,][]{2003IAUS..221P.224F, 2012ApJ...746...82F, 2017ApJ...850...71F, 2021ApJ...915...84F, 2013ApJ...768..179Y, 2017ApJ...843...61S, 2019ApJ...873...40S, 2020ApJ...902...53S}. %(e.g., Fukui et al. 2003, 2012, 2017, 2021; Yoshiike et al. 2013; Sano et al. 2017, 2019, 2020). 
These works show that the target clouds consist of neutral molecular and atomic hydrogen gas; we note that the inclusion of atomic hydrogen is indispensable since the mass of the atomic hydrogen often becomes dominant as compared {to} the molecular hydrogen. Most recently, \citet{2021ApJ...919..123S, 2021ApJ...923...15S, 2022ApJ...933..157S}
%2022arXiv220513712S} %Sano et al. (2021ab, 2022) 
presented an SNR age--$W_{\rm p}$ relation for 13 gamma-ray SNRs. They discovered a tight correlation between the SNR age and $W_{\rm p}$: the young SNRs (age \textless~6 kyrs) {show} a positive correlation, but the middle-aged SNRs (age \textgreater~8 kyrs) display a negative correlation. The authors concluded that the latter is at least caused by time-dependent diffusion (or escape) of cosmic rays from the SNR. $W_{\rm p}$ is here assumed to give a good measure of the hadronic gamma{-}rays, {although some leptonic gamma{-}ray component might be included.}
%while $W_{\rm p}$ may include some leptonic gamma ray component. 
The recent work which quantified the hadronic and leptonic gamma{-}rays in RX~J1713.7$-$3946 supports that the two contributions are indeed in the same order of magnitude \citep[e.g.,][]{2021ApJ...915...84F}.
%(Fukui et al. 2021). 
Although the age--$W_{\rm p}$ relation is crucial in understanding the acceleration and escape mechanisms of cosmic rays in SNRs, the observed samples are not large enough, especially for the middle-aged SNRs.

Puppis~A is a middle-aged SNR having a shell-like morphology with a 50 arcmin diameter in radio continuum. {This SNR is a bright thermal-Xray emitter} %thermal X-rays emitter}
%It is a bright SNR in the completely thermal X-rays
%from the shocked interstellar medium (ISM) \citep[e.g.,][]{2008ApJ...676..378H}
%(e.g., Hwang, Petre \& Flanagan 2008) 
as observed with the X-ray telescopes; Einstein \citep{1982ApJ...258...22P},
%(Petre et al. 1982)
 ROSAT (Aschenbach et al. 1993), Suzaku \citep{2008ApJ...676..378H},
%(Hwang et al. 2008)
 Chandra \citep{2005ApJ...635..355H, 2013A&A...555A...9D},
%(Hwang et al. 2005; Dubner et al. 2013)
 and XMM-Newton \citep{2006A&A...454..543H, 2010cxo..prop.3113K, 2012ApJ...756...49K, 2013A&A...555A...9D}{.}
%(Hui \& Becker 2006; Katsuda et al. 2010, 2012; Dubner et al. 2013).
{Such X-ray emission comes from the shock interstellar medium \citep[ISM; e.g.,][]{2008ApJ...676..378H}.} The compact central object (CCO) RX~J0822$-$4300 was detected in the X-rays inside the shell, and it is likely that the progenitor of the SNR {was} a high-mass star {\citep{1996ApJ...465L..43P, 1999ApJ...525..959Z, 2017MNRAS.464.3029R}}. \citet{2012ApJ...759...89H}
%Hewitt et al. (2012) 
reported the detection of faint GeV gamma-ray emission from Puppis A with the Fermi-LAT. Considering the multi-wavelength data from the radio to gamma-ray, both leptonic and hadronic models are possible with different magnetic field strengths and different energies of relativistic particles \citep[e.g.,][]{2017ApJ...843...90X}.
%(e.g., Xin et al. 2017).

The ISM interacting with the SNR is essential for understanding the origin of the high and very high energy radiation. \citet{1988A&AS...75..363D}
%Dubner \& Arnal (1988) 
observed the {H{\sc i}} 21 cm line and the CO($J$ = 1--0) 2.6 mm line and found a molecular cloud coincident with the SNR shell. These authors interpreted that this coincidence shows an interacting cloud at $\sim$16~km~s$^{-1}$. {An} H{\sc i} study with VLA supported the velocity based on morphological coincidence \citep{1995AJ....110..318R, 2003MNRAS.345..671R},
%(Reynoso et al. 1995; 2003)
 and these works suggested a kinematic distance of 2.2~kpc. Follow up studies of CO or OH lines however did not give conclusive results on the interacting clouds by non-detections due to too small spatial coverage/low resolution {(CO, \citeauthor{2008A&A...480..439P} \citeyear{2008A&A...480..439P}; 1720-MHz OH, \citeauthor{1996AJ....111.1651F} \citeyear{1996AJ....111.1651F}; four lines of OH,  \citeauthor{2000MNRAS.317..421W} \citeyear{2000MNRAS.317..421W}).}
%\citep[CO,][]{2008A&A...480..439P}, % (CO,\citet{2008A&A...480..439P}%Paron et al. 2008; 1720-MHz OH, 
%\citep[1720-MHz OH,][]{1996AJ....111.1651F}%\citet{1996AJ....111.1651F}%Frail et al. 1996; four lines of OH 
%\citep[four lines of OH,][]{2000MNRAS.317..421W}. %\citet{2000MNRAS.317..421W}%Woermann et al. 2000). %referencesの入れ方確認
% \citeauthor{2005ASPC..342..105N} \citeyear{2005ASPC..342..105N}
Subsequently, \citet{2017MNRAS.464.3029R} %Reynoso et al. (2017) 
conducted new H{\sc i} absorption measurements with ATCA, and found that $\sim$1.3~kpc is a more likely distance, {in contrast to} %which contradicts 
their early results. In summary, more extensive efforts are desirable in order to have a comprehensive physical picture of the ISM in Puppis A and better constrain its distance.

In the present work, we have undertaken a new study of the ISM toward Puppis~A by employing the CO and  H{\sc i} data which are combined with the X-rays and the gamma{-}rays as well as the visual extinction. The present paper is organized as follows. Section \ref{observations} described the datasets and Section \ref{results} presents the results, including the distributions of CO and H{\sc i} intensity and their kinematic properties. Section \ref{discussion} discusses the ISM associated with the SNR and that located in front of the SNR along with the relationship with the X-rays and gamma{-}rays. New pieces of evidence for the associated ISM are presented and a distance is established. Section \ref{conclusions} concludes the paper.

\section{OBSERVATIONS AND DATA REDUCTION}\label{observations}
\subsection{CO}\label{observations:co}
Observations of $^{12}$CO($J$~=~2--1) at 230.538~GHz were conducted from 2014 December to 2015 January using the NANTEN2 4-m millimeter/sub-millimeter telescope at Pampa la Bola in northern Chile (4,865 m above sea level). We used the on-the-fly (OTF) mode with Nyquist sampling, and the observed area was 75$'$ $\times$ 30$'$. The front end was a 4-K cooled Nb superconductor-insulator-superconductor (SIS) mixer receiver. The typical system temperature including the atmosphere was $\sim$200~$\pm$~40~K in the double sideband. The back end was a digital Fourier-transform spectrometer (DFS) with 1~GHz bandwidth and 
%1684 
{16384} channels, corresponding to a velocity resolution of 0.08~km s$^{-1}$ 
%.
and a velocity coverage of 1300~km~s$^{-1}$. The pointing offset was better than $\sim$10$''$, verified by observing Jupiter and IRC+10216. The absolute intensity was calibrated by observing Orion-KL [$\alpha(\mathrm{J2000}) = 05^\mathrm{h}35^\mathrm{m}13\fs471, \delta(\mathrm{J2000}) = -5\arcdeg22\arcmin27\farcs55$] \citep{2014ApJ...795...13B}%(Bern\UTF{00E9} et al. 2014)
, obtaining a main beam-efficiency %is
$\sim$0.63. After convolution using a three-dimensional Gaussian function of 60$''$ (FWHM), the final beam size was $\sim$100$''$. The typical noise fluctuations are $\sim$0.14~K at a velocity resolution of 1~km~s$^{-1}$.

In addition, we used the $^{12}$CO($J$~=~1--0) emission line data at 115.271~GHz taken with the NANTEN 4-m telescope, which were already published in \citet{2001PASJ...53.1025M}. %Moriguchi et al. (2001). 
Observations were carried out in the position-switching mode with a 2$'$ grid spacing. The angular resolution of the data was 2.6$'$ (FWHM) and the velocity resolution was 0.65~km~s$^{-1}$. The typical noise fluctuations are $\sim$0.16~K at a velocity resolution of 1~km~s$^{-1}$ \citep[for more detailed information, see][]{2001PASJ...53.1025M}. %(for more detailed information, see \citet{2001PASJ...53.1025M}).%Moriguchi et al. 2001).
%引用の仕方確認

\begin{figure*}[]
\begin{center}
\includegraphics[width=\linewidth,clip]{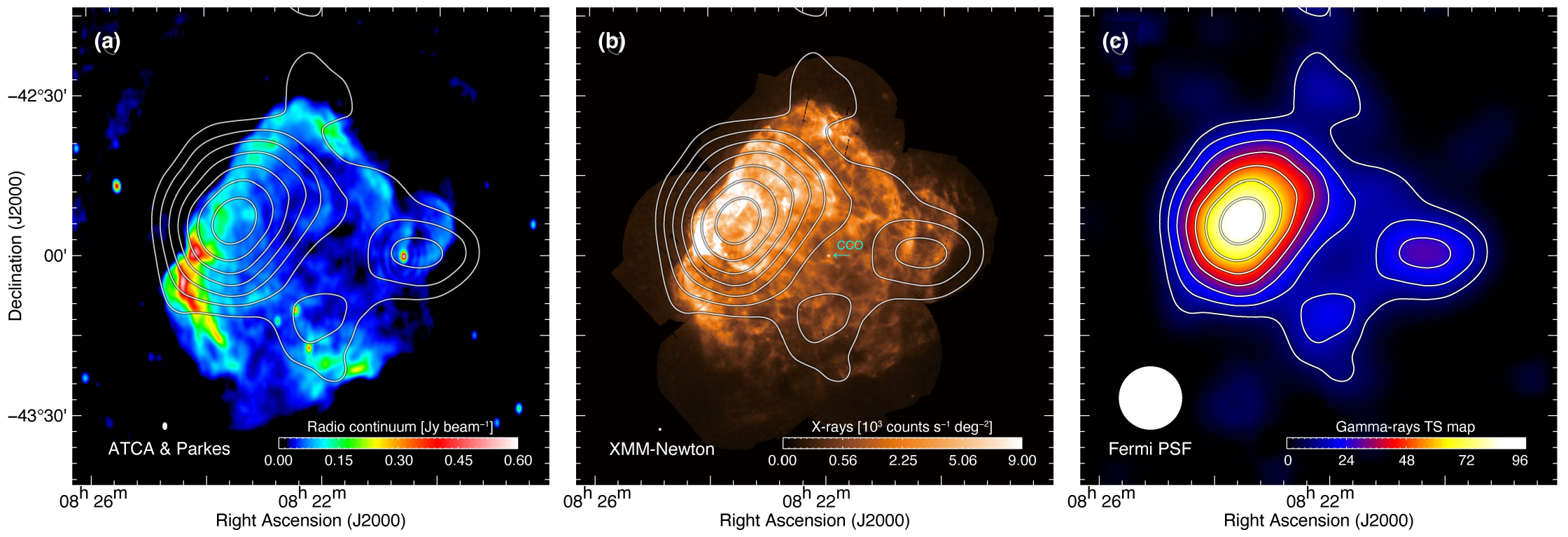}
\caption{Maps of (a) 1.4 GHz radio continuum from the ATCA $\&$ Parkes \citep{2017MNRAS.464.3029R}, (b) the XMM-Newton $\&$ Chandra X-ray flux ($E$: 0.5--7.0 keV), and (c) gamma-rays above 10GeV toward Puppis A \citep{2017ApJ...843...90X}. The superposed contours are the gamma-rays and the contour levels are 3, 4, 5, 6, 7, 8, and 9 sigma.}
\label{fig1}
\end{center}
\end{figure*}%

\subsection{H{\sc i}}\label{observations:hi}
We used the 21~cm H{\sc i} data taken with the Australia Telescope Compact Array (ATCA) and single-dish data from the Parkes 64~m radio telescope{.} {Details on the observing techniques and data processing are given in \citet{2017MNRAS.464.3029R}.} %Reynoso et al. 2017}. 
The beam size of H{\sc i} was 118.3$''$ $\times$ 88.9$''$, with a position angle of $-4.3^{\circ}$. The typical noise level is $\sim$0.71~K at a velocity resolution of 1~km~s$^{-1}$.

\subsection{X-rays}\label{observations:xrays}
To compare the spatial distributions of CO/H{\sc i} and X-rays in detail, we analyzed archival X-ray data obtained by XMM-Newton and Chandra \citep[e.g.,][]{2010cxo..prop.3113K, 2012ApJ...756...49K, 2013ApJ...768..182K, 2013A&A...555A...9D, 2016A&A...590A..70L}.%(e.g., Katsuda et al. 2010, 2012, 2013; Dubner et al. 2013; Luna et al. 2016).
%引用の仕方確認

Archival XMM-Newton data taken with both the EPIC-pn and EPIC-MOS were reduced using the XMM-Newton Science Analysis System \citep[SAS;][]{2004ASPC..314..759G} %(SAS; \citet{2004ASPC..314..759G})%Gabriel et al. 2004)
 version 19.1.0 and HEAsoft version 6.28. We reprocessed the observation data files following the procedure provided {as} the XMM-Newton extended source analysis software \citep[ESAS;][]{2008A&A...478..575K}, %(ESAS; \citet{2008A&A...478..575K}%Kuntz \& Snowden 2008)
 except for the EPIC-pn data taken in the large-window mode. After filtering soft proton flares using {“mos-/pn-filter”} %“mos-/~pn-~filter” 
 tasks, we generated quiescent particle background (QPB) images and exposure maps for each observation using the tasks of {“mos-/pn-spectra”} %“mos-/~pn-spectra” 
and {“mos-/pn-back”.} %“mos-/pn-back.” 
We also used the “eimageget” task for the EPIC-pn data taken in the large-window mode. The “merge\_comp\_xmm” task was used to combine all the XMM-Newton data. Here we created the background, exposure, and counts maps for the energy bands of 0.5--7.0~keV, 0.36--0.46~keV, and 1.14--1.27~keV. The former energy band represents the broadband image, and the others correspond to the continuum bands without any strong line emission \citep[cf.][]{2008ApJ...676..378H}.%(cf. Hwang et al. 2008).
%"****", /~pn
%引用の仕方確認

%\clearpage
\begin{figure*}[]
\begin{center}
\includegraphics[width=150mm,clip]{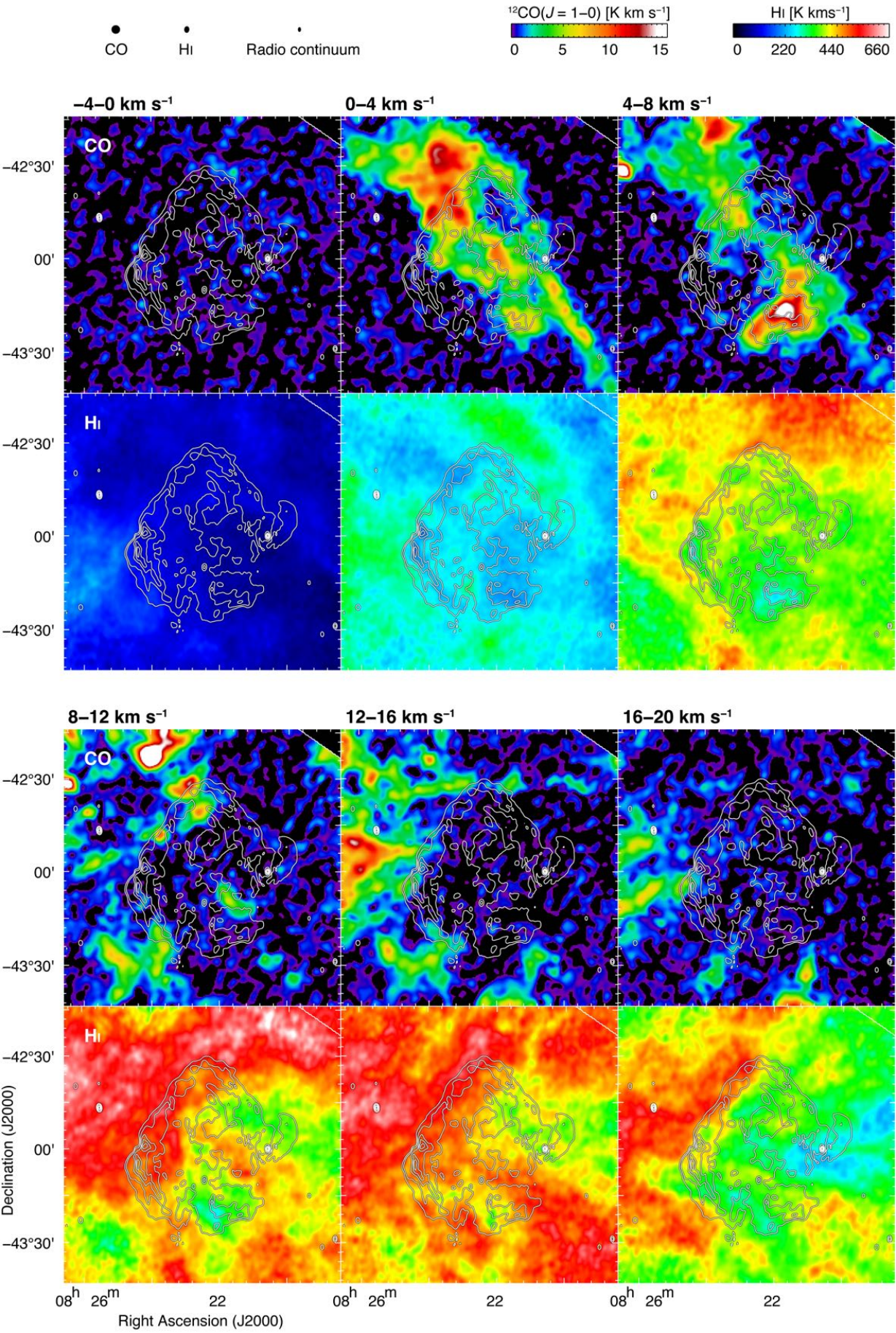}
\caption{{Velocity channel distributions of the NANTEN  $^{12}$CO($J$ = 1--0) and  the ATCA $\&$ Parkes H{\sc i}. The superposed contours are 1.4 GHz radio continuum from the ATCA $\&$ Parkes \citep{2017MNRAS.464.3029R}. The contour levels are 40, 100, 220, 340, and 460 mJy beam$^{-1}$. Each panel of CO/H{\sc i} shows distributions every 4 km s$^{-1}$ in a velocity range from $-$4 to 20 km s$^{-1}$. The %scale and 
color bars for CO and H{\sc i} are shown on top of the set of panels.}}
\label{fig2}
\end{center}
\end{figure*}%

In the Chandra analysis, we used nine individual observational datasets that were taken with the Advanced CCD Imaging Spectrometer (ACIS). All the datasets were reduced using the Chandra Interactive Analysis of Observations \citep[CIAO,][]{ 2006SPIE.6270E..1VF} %(CIAO, \citet{ 2006SPIE.6270E..1VF}) %Fruscione et al. 2006)
software version 4.12 with CALDB 4.9.1 \citep{2007ChNew..14...33G}. %(Graessle et al. 2007).
We utilized the “chandra\_repro” task to reprocess the data with the latest calibration. We then created the background, exposure, and counts maps for each energy band using the “fluximage” task.

To combine both the XMM-Newton and Chandra datasets, we corrected vignetting and weighting according to their respective effective areas for each energy band. After applying the adaptive smoothing using the “adapt\_merge” task in ESAS, we then finally obtained exposure-corrected, energy-filtered, and background-subtracted images covering the entire SNR.

\subsection{Other wavelength datasets}\label{observations:other wavelength datasets}
H$\alpha$ and radio continuum data are used to derive the spatial distribution of the ionized gas and synchrotron radiation, respectively. We used the H$\alpha$ data that appear in the Super Cosmos H$\alpha$ Survey \citep[SHS,][]{2005MNRAS.362..689P}, %(SHS; Parker et al. 2005)
and the 1.4~GHz radio continuum data taken with ATCA and the Parkes 64~m radio telescope \citep{2017MNRAS.464.3029R}. %(Reynoso et al. 2017). 
The angular resolution is 0.67$''$ for the SHS H$\alpha$ data and 82.2$''$ $\times$ 50.6$''$ with a position angle of $-0.55^{\circ}$ for the radio continuum data.

\section{Results}\label{results}
\subsection{Distributions of Gamma-ray, Radio continuum, and X-ray}\label{results:overview}
%Overview of X-ray and H{\sc i} Distributions, Distributions of X-ray and Radio continuum, Distributions of Gamma-ray, Radio continuum, and X-ray
Figure \ref{fig1}(a) shows the distribution of the 1.4~GHz radio continuum obtained from the ATCA \& Parkes \citep{2017MNRAS.464.3029R}. %(Reynoso et al. 2017). 
The radio continuum shows a shell, which is bright and flat on the northeastern side. The shell also has additional components in the west at Dec. $\sim$$-43\arcdeg$, R.A. $\sim$$8^\mathrm{h} {20}^\mathrm{m}$ %{20}^\mathrm{m}$
and in the south Dec. $\sim$$-43\arcdeg25\arcmin$, R.A. $\sim$$8^\mathrm{h}22^\mathrm{m}{48^\mathrm{s}}$, which are not continuous with the shell.

Figure \ref{fig1}(b) shows the composite X-ray image of Puppis A in the energy band 0.5--7.0~keV \citep[e.g.,][]{2008ApJ...678..297K, 2013A&A...555A...9D}. %(e.g., Katsuda et al. 2008, Dubner et al. 2013). 
%The X-rays are completely thermal and are center-filled with filamentary distribution. 
{The X-rays are completely thermal and are distributed nearly within the boundaries set by the radio emission shell, with an intensity gradient
decreasing from the bright eastern edge to the west. A broad strip of harder X-ray emission covers the center of the SNR in the NE-SW direction.}
{There is a central compact object (CCO) in the center of the SNR.} %as marked by a cross. 
In addition, X-rays are bright in the northeastern half and have good correspondence with the radio continuum radiation.

Figure \ref{fig1}(c) shows the GeV gamma{-}ray distribution obtained with the Fermi-LAT \citep{2017ApJ...843...90X}. %(Xin et al. 2017). 
The major peak coincides with the X-ray shell and the two additional peaks in the west and south seem to correspond to the radio and X-ray components (Figures \ref{fig1}a and \ref{fig1}b). %(Figures 1a and 1b).
%GeV gamma ray, 

\subsection{Velocity channel distributions of CO and H{\sc i}}\label{results:channel map}
Figure \ref{fig2} upper panel shows the {integrated velocity channel maps of the NANTEN $^{12}$CO($J$~=~1--0) every 4~km~s$^{-1}$} from $-$4~km~s$^{-1}$ to 20~km~s$^{-1}$. We find the largest CO cloud in the field is located at $V_{\rm LSR}$ = 0--8~km~s$^{-1}$ and is elongated from the northeast to the southwest over $\sim$1~degree. Several smaller CO clouds are distributed toward the SNR shell at  $V_{\rm LSR}$ = 0--20~km~s$^{-1}$.

%\clearpage
\begin{figure*}[]
\begin{center}
\includegraphics[width=\linewidth]{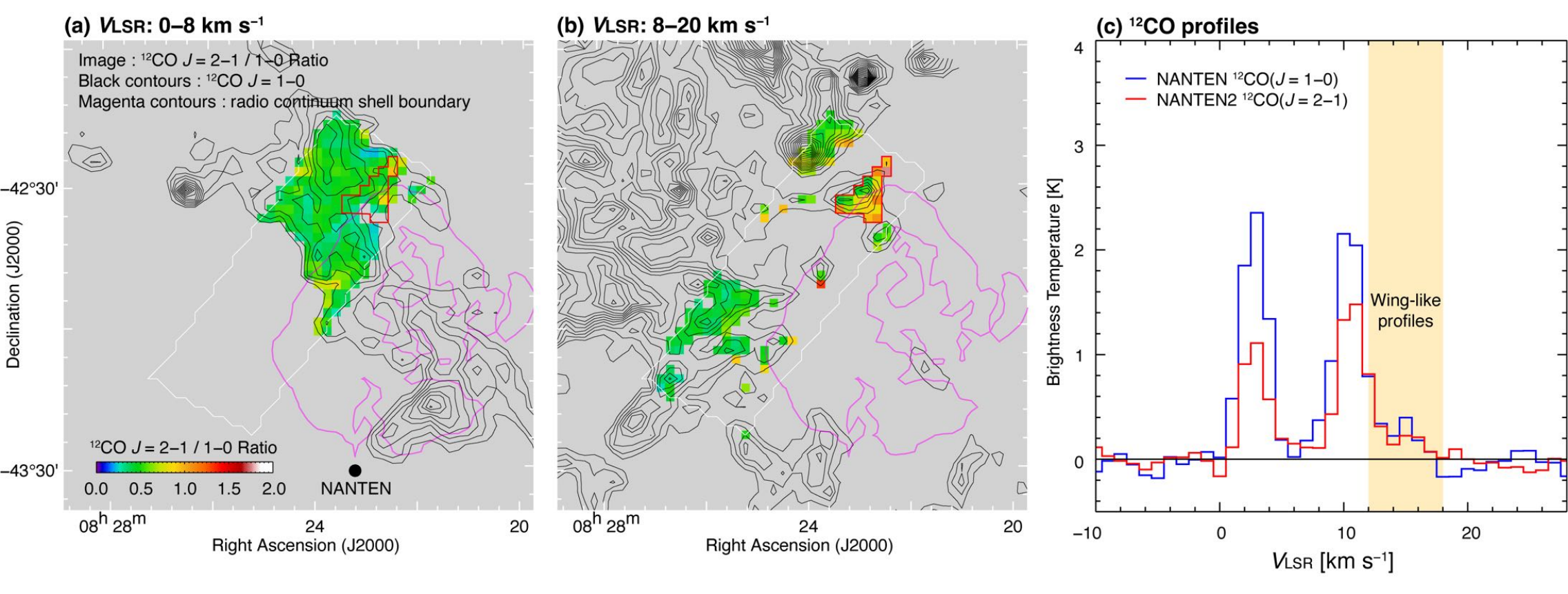}
\vspace*{-0.8cm}
\caption{(a--b) Maps of $^{12}$CO($J$ = 2--1/ 1--0)  ratios at the two velocity clouds, superposed on $^{12}$CO($J$ = 1--0) contours (black) and radio continuum shell boundary (magenta). The lowest contour level and the contour intervals for CO are 3 and 2 K km s$^{-1}$, respectively. The contour level of the radio continuum shell boundary is 40 mJy beam$^{-1}$. The $^{12}$CO($J$ = 2--1) data were smoothed to match the effective beam size of the $^{12}$CO($J$ = 1--0) data (angular resolution $\sim$156$''$). The gray areas represent that the $^{12}$CO($J$ = 1--0)  and/or $^{12}$CO($J$ = 2--1) data show a low significance of $\sim$5$\sigma$ or lower. The white rectangles indicate the observed areas of $^{12}$CO($J$ = 2--1) . (c) Averaged $^{12}$CO($J$ = 2--1)  (red) and $^{12}$CO($J$ = 1--0) (blue) profiles which are extracted from red rectangles as shown in Figures \ref{fig3}a and \ref{fig3}b. The yellow shaded area represents the velocity range of wing-like profiles (see the text).}
%\red{The superposed contours in the bottom panels indicate $^{12}$CO($J$ = \red{1--0}), whose contour levels are ******, and **** for Figure \ref{fig4}g; ******, and **** for Figure \ref{fig4}h; and ******, and **** for Figure \ref{fig4}i.} The dashed arches in the p--v diagrams indicate the boundaries of the CO and H{\sc i} cavities (see the text).}
\label{fig3}
\end{center}
\end{figure*}%

\begin{figure*}[]
\begin{center}
\includegraphics[width=102mm,clip]{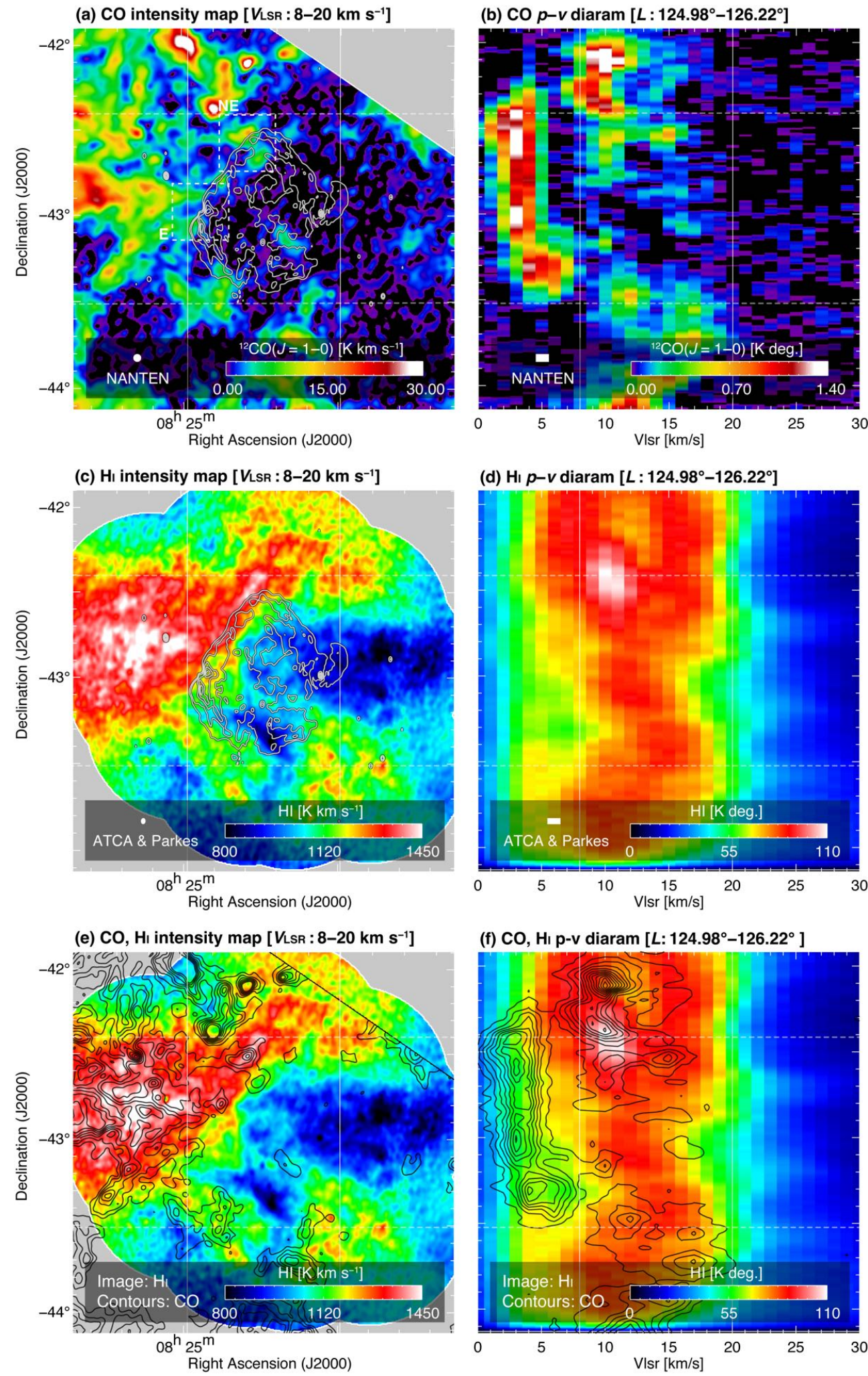}
\caption{Integrated intensity maps and position--velocity (p--v) diagrams of $^{12}$CO($J$ = 1--0) (top panels) and H{\sc i} (middle and bottom panels). The integration range is  8 to 20 km s$^{-1}$ (left panels) in the velocity for each intensity map; and from 124\fdg98 to 126\fdg22 in the right ascension for each p--v diagram. The superposed contours in Figures \ref{fig4}a and \ref{fig4}c indicate the radio continuum at 1.4 GHz whose contour levels are the same as shown in Figure \ref{fig2}. The superposed contours in the bottom panels indicate $^{12}$CO($J$ = 1--0), whose contour levels are 3, 6, 9, 12, 15, 18, and 21 K km s$^{-1}$ for Figure \ref{fig4}e; and 0.10, 0.28, 0.46, 0.64, 0.82, 1.00, 1.18, 1.36, 1.54, 1.72, 1.90, and 2.08 K degree for Figure \ref{fig4}f. }
%The dashed arches in the p--v diagrams indicate the boundaries of the CO and H{\sc i} cavities (see the text).}
%白線取り除いたので上記一文削除。
%\caption{Integrated intensity maps and position--velocity (p--v) diagrams of $^{12}$CO($J$ = 1--0) (top panels) and H{\sc i} (bottom panels). The integration range is from 0 to 8 km s$^{-1}$ (left panels) and 8 to 20 km s$^{-1}$ (middle panels) in the velocity for each intensity map; and from 124\fdg98 to 126\fdg22 in the right ascension for each p--v diagram. The superposed contours in Figures \ref{fig4}a, \ref{fig4}b, \ref{fig4}d, and \ref{fig4}e indicate the radio continuum at 1.4 GHz whose contour levels are the same as shown in Figure \ref{fig2}. The dashed arches in the p--v diagrams indicate the boundaries of the CO and H{\sc i} cavities (see the text).}
\label{fig4}
\end{center}
\end{figure*}

Figure \ref{fig2} lower panel shows the velocity channel distributions of the ATCA \& Parkes H{\sc i} in the same velocity ranges with Figure \ref{fig2} upper panel. The H{\sc i} emission is brighter than 400~K~km~s$^{-1}$ at  $V_{\rm LSR}$  = 8--16~km~s$^{-1}$, while the H{\sc i} intensity at  $V_{\rm LSR}$  = 0--8~km~s$^{-1}$ is significantly weaker toward the radio shell of the SNR. The H{\sc i} clouds at  $V_{\rm LSR}$ = 8--20~km~s$^{-1}$ is distributed along the eastern half of the SNR shell, and the edge of the H{\sc i} clouds at  $V_{\rm LSR}$  = 16--20~km~s$^{-1}$ shows a good spatial correspondence with the northeastern shell of the SNR. 
%The 0--8~km~s$^{-1}$ cloud and the 8--20~km~s$^{-1}$ cloud are hereafter called “the 3~km~s$^{-1}$ cloud” and “the {11}%16~km~s$^{-1}$ cloud”, respectively, following the convention in the previous papers. 
More detailed comparison of the clouds with the shell is shown later in Figure \ref{fig4}.

%\clearpage

\subsection{The CO 2--1/1--0 ratio}\label{results:ratio map}
Figures \ref{fig3}(a) and \ref{fig3}(b) show maps of $^{12}$CO~$J$~=~2--1/1--0 intensity ratio (hereafter $R_{\rm CO}$) in the two velocity ranges from 0~km~s$^{-1}$ to 8~km~s$^{-1}$ and from 8~km~s$^{-1}$ to 20~km~s$^{-1}$. The $J$~=~1--0 mapping is limited to the rectangular area shown by white lines which includes the northeastern part of the shell. The ratio between the two transitions is convolved to the beam size of the $J$ = {2--1} transition. {The CO cloud at 3~km~s$^{-1}$} %The 3~km~s$^{-1}$ CO cloud 
shows a low $R_{\rm CO}$ value of $\sim$0.5 in the entire cloud, whereas the %16{11}~km~s$^{-1}$ CO cloud 
{cloud at 11~km~s$^{-1}$} shows a higher $R_{\rm CO}$ value %up to $\sim$1.8 in the northeast. 
 $\sim$0.8--1.1{.} Figure \ref{fig3}(c) shows averaged $^{12}$CO($J$~=~2--1) and $^{12}$CO($J$~=~1--0) profiles enclosed by the red rectangular area in Figures \ref{fig3}(a) and \ref{fig3}(b). 
%{We find wing-like profiles in the CO emission lines at $V_{\rm LSR}$ =  {12--18}~km~s$^{-1}$. Since the typical noise fluctuation of averaged profile is $\sim$0.06~K/ch for $^{12}$CO($J$~=~2--1) emission line, the wing-like profile has a signal-to-noise ratio of 5 or greater.}
{We find wing-like profiles in both of the CO profiles at $V_{\rm LSR}$ =  {12--18}~km~s$^{-1}$. }
% {We find wing-like profile in the CO profile at $V_{\rm LSR}$ =  {12--18}~km~s$^{-1}$.}
% {Since the typical noise fluctuation of averaged profile is $\sim$0.06~K/ch for the $^{12}$CO($J$~=~2--1) emission lines, the wing-like profile has signal-to-noise ratios~$>$~5.}
% {Since the typical noise fluctuations of averaged profiles are $\sim$0.03~K/ch for both the $^{12}$CO($J$~=~1--0) and $^{12}$CO($J$~=~2--1) emission lines, the wing-like profiles have signal-to-noise ratios~$>$~5.}
{Since the typical noise fluctuations of averaged profiles are $\sim$0.07~K/ch for the $^{12}$CO($J$~=~1--0) and $\sim$0.06~K/ch for the $^{12}$CO($J$~=~2--1) emission lines, the wing-like profiles have signal-to-noise ratios~$>$~4.}
% {The $T_{\mathrm rms}$ of averaged profiles of $^{12}$CO($J$~=~2--1) and $^{12}$CO($J$~=~1--0) are $\sim$0.03K, respectively, and the brightness temperature of wing-like profiles are greater than $\sim$5 sigma.}
 The ratio distribution indicates that the %16
{cloud at 11~km~s$^{-1}$} % {11}~km~s$^{-1}$ cloud 
is more highly excited than the %3~km~s$^{-1}$ cloud
{cloud at 3~km~s$^{-1}$}. The ratio more than {$\sim$}1 is typical {of}
 %to 
 the cloud shocked by SNRs, while the ratio $\sim$0.5 is typical {of}
 %to 
 non-shocked gas \citep[e.g., for W44 see][]{2013ApJ...768..179Y}. %(e.g., for W44 see \citet{2013ApJ...768..179Y}). %Yoshiike et al. 2013). 
 Along with the broad CO wing, the high line intensity ratio CO indicates that the %{11}%16 ~km~s$^{-1}$ cloud 
 {cloud at 11~km~s$^{-1}$} only is interacting with the SNR.
 {H{\sc i} or molecular features below 8~km~s$^{-1}$ are hardly related to Puppis A, since the previous absorption studies place the SNR well beyond this velocity \citep[e.g.,][]{2000MNRAS.317..421W, 2017MNRAS.464.3029R}. %(e.g.Woermann et al. 2000, Reynoso et al. 2017). 
Besides, the peak at 3~km~s$^{-1}$ shown in Figure \ref{fig3}(c) is narrow, as expected for undisturbed gas. 
 In all, this component is most likely foreground to the SNR, hence we will focus on the 8--20~km~s$^{-1}$ cloud.}

\subsection{Detailed comparison of CO and H{\sc i} with the SNR shell}\label{results:pv map}
We focus on the %16
{8--20}~km~s$^{-1}$ cloud and compare it with the SNR shell into more detail. Figures \ref{fig4}(a), \ref{fig4}(c), and \ref{fig4}(e) show the CO and H{\sc i} in the velocity range 8--20~km~s$^{-1}$, which are superposed on the radio continuum, respectively, and are superposed with each other. Figures \ref{fig4}(b), \ref{fig4}(d), and \ref{fig4}(f) show p--v diagrams of the %16
{8--20}~km~s$^{-1}$ cloud in CO and H{\sc i}. 
{The p--v diagrams at 8--20~km~s$^{-1}$ show a cavity for $-42\fdg5$--$-42\fdg4$ which corresponds to the SNR shell.}
%The curved lines in the p--v diagrams show the loci of {a cavity} %an expanding shell %for radii of 0.55$^{\circ}$ and 0.25$^{\circ}$ with expansion velocity of $\sim$12~km~s$^{-1}$ and $\sim$5~km~s$^{-1}$. 
%The densest part of the shell is perhaps traced in CO, and is expanding at the same velocity and radius as the H{\sc i}.
% The curved lines in the p--v diagrams show the loci of an expanding shell for radii of 0.55$^{\circ}$ and 0.25$^{\circ}$ with expansion velocity of $\sim$12~km~s$^{-1}$ and $\sim$5~km~s$^{-1}$.

\begin{figure*}[]
\begin{center}
\includegraphics[width=\linewidth]{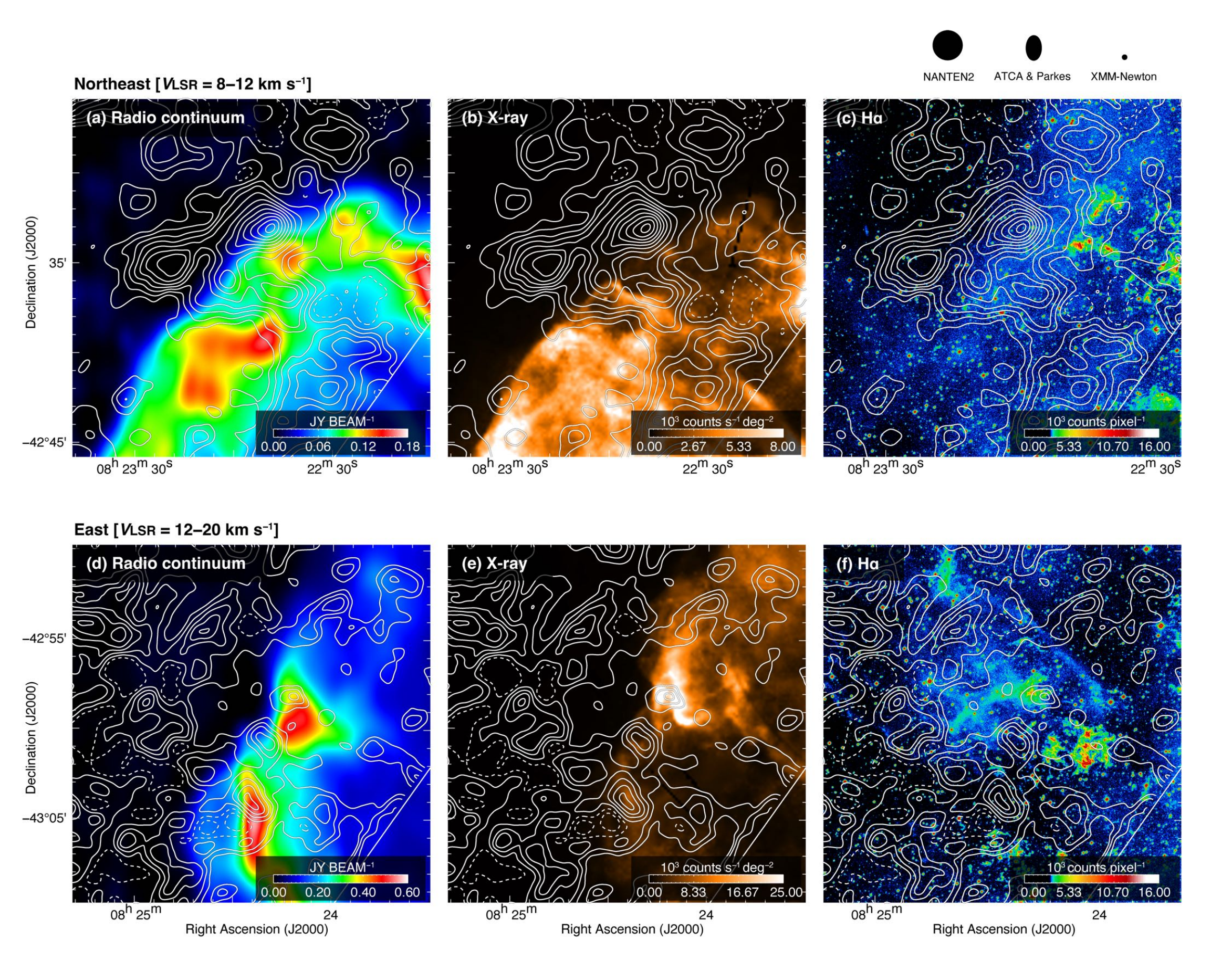}
\vspace*{-1.cm}
\caption{{Enlarged views of northeastern shell (top panels), and the eastern shell (bottom panels) as indicated by the dashed white rectangles  in Figure \ref{fig4}a. 
{Distribution of the $^{12}$CO($J$ = 2--1) contours superposed on the radio continuum images (left panels), X-ray images (middle panels), and H$\alpha$ images \citep[right panels,][]{2005MNRAS.362..689P}.}
%Distribution of the radio continuum images (left panels), X-ray images (middle panels), and H$\alpha$ images \citep[right panels,][]{2005MNRAS.362..689P} superposed on the $^{12}$CO($J$ = 2--1) contours. 
The lowest contour level and the contour intervals are 1.0 and 1.0 K km s$^{-1}$, respectively. The integration velocity range is 8--12 km s$^{-1}$ for the top panels and 12--20 km s$^{-1}$ for the bottom panels.}}
%(right panels\red{, ref.}) の箇所は、\citep[right panels,][]{******} として、******の箇所にH$\alpha$サーベイのリファレンスを入れればOKです (HS)。
\label{fig5}
\end{center}
\end{figure*}

We compare spatial distributions among CO, X-rays, radio continuum, and H$\alpha$ toward the two areas in the northeast and in the east of the shell as indicated in Figure \ref{fig4}a. The upper panels of Figure \ref{fig5}, for the northeast of the shell, compare the $^{12}$CO($J$~=~2--1) contours with the radio continuum image (Figure \ref{fig5}a), the X-ray image (Figure \ref{fig5}b), and the H$\alpha$ image (Figure \ref{fig5}c). The lower panels of Figure \ref{fig5}, for the east of the shell, compare the $^{12}$CO($J$~=~2--1) contours with the radio continuum image (Figure \ref{fig5}d), the X-ray image (Figure \ref{fig5}e), and the H$\alpha$ image (Figure \ref{fig5}f).

We find that the X-rays at Dec.~$\textless$~$-42\arcdeg39\arcmin$ and %R.A.~$\textgreater$~$8^\mathrm{h}23^\mathrm{m}50^\mathrm{s}$
{R.A.~$\textgreater$~$8^\mathrm{h}{22}^\mathrm{m}50^\mathrm{s}$}, tend to be anti-correlated with the CO ({Figure \ref{fig5}b}), and that the H$\alpha$ at Dec.~=~$-42\arcdeg35\arcmin$ to 30$\arcmin$ and {R.A.}
$\textless$~$8^\mathrm{h}22^\mathrm{m}50^\mathrm{s}$ tends to be anti-correlated with the CO (Figure \ref{fig5}c). Similar trends are recognized in the radio continuum image compared with the CO (Figure \ref{fig5}a). Further, we find that X-ray features and H$\alpha$ features are located within the small CO cavity at Dec.~=~$43\arcdeg00\arcmin$ to $42\arcdeg55\arcmin$ and at %R.A.~=~$8^\mathrm{h}43^\mathrm{m}50^\mathrm{s}$ 
R.A.~=~$8^\mathrm{h}{23}^\mathrm{m}50^\mathrm{s}$ 
to $8^\mathrm{h}24^\mathrm{m}40^\mathrm{s}$ (Figures \ref{fig5}{e} %d 
and \ref{fig5}{f}%e
). These X-ray features were suggested to be interacting with the ISM previously by \citet{2005ApJ...635..355H} %Hwang et al. (2005) 
without a direct comparison with the ISM. The present result (Figure \ref{fig5}e) indicates a small CO clump of 1~arcmin size toward the X-rays, which shows a possible interaction candidate. To summarize, the intense parts of the SNR shell show a trend that the %16
{8--20}~km~s$^{-1}$ CO cloud is {anti-correlated} with the ionized or hot gas/high energy electrons, lending support for the association of the %16
{8--20}~km~s$^{-1}$ cloud with the SNR.

%\clearpage

\section{Discussion}\label{discussion}
\subsection{Molecular and Atomic Clouds Associated with the SNR Puppis~A}\label{discussion:associated}
\subsubsection{Comparison with the X-ray hardness ratio and $A_{V}$}\label{results:Hardness ratio Av map}
In order to test the interstellar absorption of the X{-}rays, we compare the hardness ratio map of X-rays with the total interstellar proton column density $N_{\rm p}$(H$_{2}$ + H{\sc i}) of the {0--8~km~s$^{-1}$} %3~km~s$^{-1}$ 
and %16
{8--20~km~s$^{-1}$} %{11}~km~s$^{-1}$ 
clouds. We use the following equations to derive $N_{\rm p}$(H$_{2}$ + H{\sc i}) for each cloud:
\begin{eqnarray}
N_{\mathrm p}(\mathrm H_{2} + \mathrm {H}{\textsc{i}}) = N_\mathrm{p}(\mathrm H_{2}) + N_\mathrm{p}(\mathrm{H}{\textsc{i}}).
\label{eq1}
\end{eqnarray}
where $N_\mathrm{p}(\mathrm H_{2})$ is the proton column density of the molecular component and  $N_{\mathrm p}(\mathrm{H}{\textsc{i}})$ is the proton
column density of the atomic component.

$N_\mathrm{p}(\mathrm H_{2})$ can be derived from the following relationships between the molecular hydrogen column
density $N(\mathrm H_{2})$ and the $^{12}$CO($J$~=~1--0) integrated intensity $W$(CO): 
\begin{eqnarray}
N(\mathrm H_{2}) = X \cdot W(\mathrm{CO}) (\mathrm{cm}^{-2}),
\label{eq2}
\end{eqnarray}
\begin{eqnarray}
N_{\mathrm p}(\mathrm H_{2}) = 2 \times N(\mathrm H_{2}) (\mathrm{cm}^{-2}).
\label{eq3}
\end{eqnarray}
where $X$ is a CO-to-H$_{2}$ conversion factor between the $N(\mathrm H_{2})$ and $W$(CO). We use $X$ = 1.5 $\times$ 10$^{20}$ cm$^{-2}$
(K~km~s$^{-1}$)$^{-1}$ which is derived in the Appendix A.

For calculating the H{\sc i} column density $N_{\mathrm p}(\mathrm{H}{\textsc{i}})$, we used a conversion factor between the optical-depth-corrected $N_{\mathrm p}(\mathrm{H}{\textsc{i}})$ and $W(\mathrm{H}{\textsc{i}})$ \citep{2015ApJ...798....6F, 2017ApJ...850...71F}%(Fukui et al. 2015; 2017)
, and obtained the average optical-depth-corrected  $N_{\mathrm p}(\mathrm{H}{\textsc{i}})$ to be $\sim${(4.3~$\pm$~0.4)} %(4.6~$\pm$~0.5) 
$\times$ 10$^{21}$~cm$^{-2}$ in the Puppis~A region, which is $\sim${2.2} %2.4 
times greater than the optically thin case.

%\clearpage
\begin{figure*}[]
\begin{center}
\includegraphics[width=\linewidth]{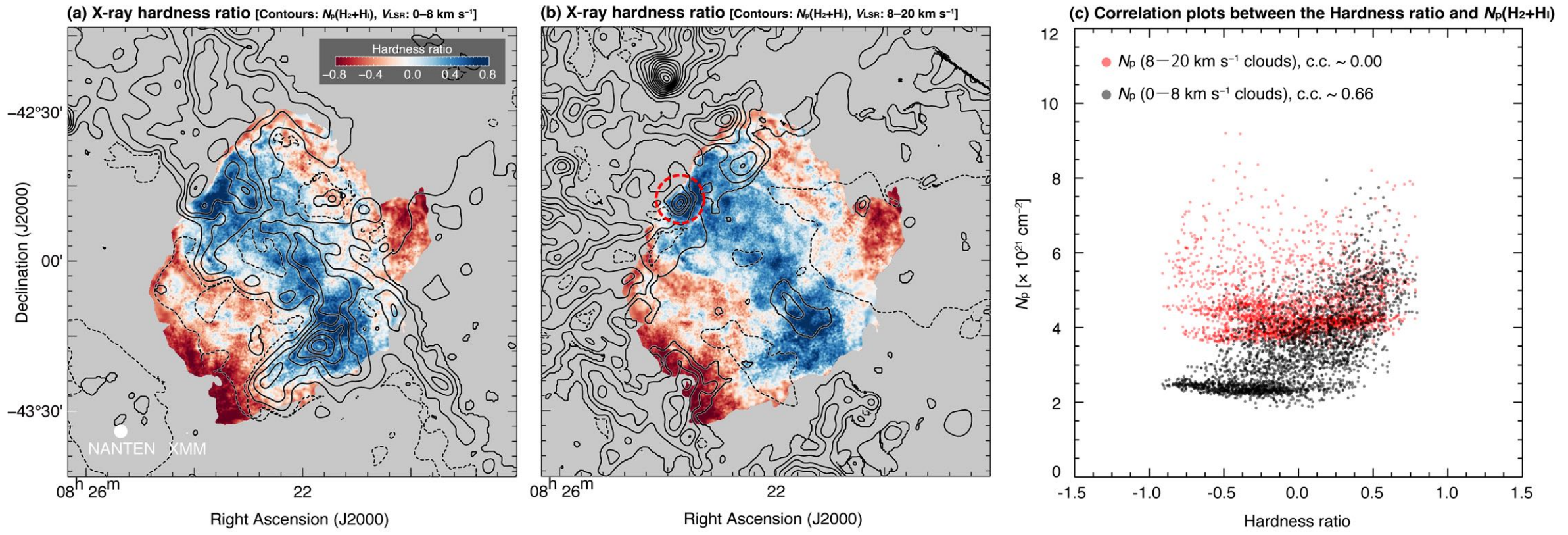}
\caption{(a--b) X-ray hardness ratio maps overlaid with the CO/H{\sc i} derived $N_{\rm p}$(H$_{2}$ $+$ H{\sc i}) contours at $V_{\rm LSR}$ = 0--8 km s$^{-1}$ (left panel) and 8--20 km s$^{-1}$(middle panel). The lowest contour level and the contour intervals are 0.8 and 0.8 $\times 10^{21}$cm$^{-2}$, respectively. The energy band of the hardness ratio map is 0.36--0.46 keV for the soft-band image and 1.14--1.27 keV for the hard band image. The gray areas are excluded regions outside the radio continuum shell.  {The dashed circle in (b) encloses a local spot with very high hardness ratio.} (c) Correlation plots between the Hardness ratio and CO/H{\sc i} derived $N_{\rm p}$(H$_{2}$ $+$ H{\sc i}). The black and red filled circles represent the data points in the velocity ranges of 0--8 km s$^{-1}$, and 8--20 km s$^{-1}$, respectively.}
%(greater than 3sigma) 
\label{fig6}
\end{center}
\end{figure*}%

\begin{figure*}[]
\begin{center}
\includegraphics[width=\linewidth]{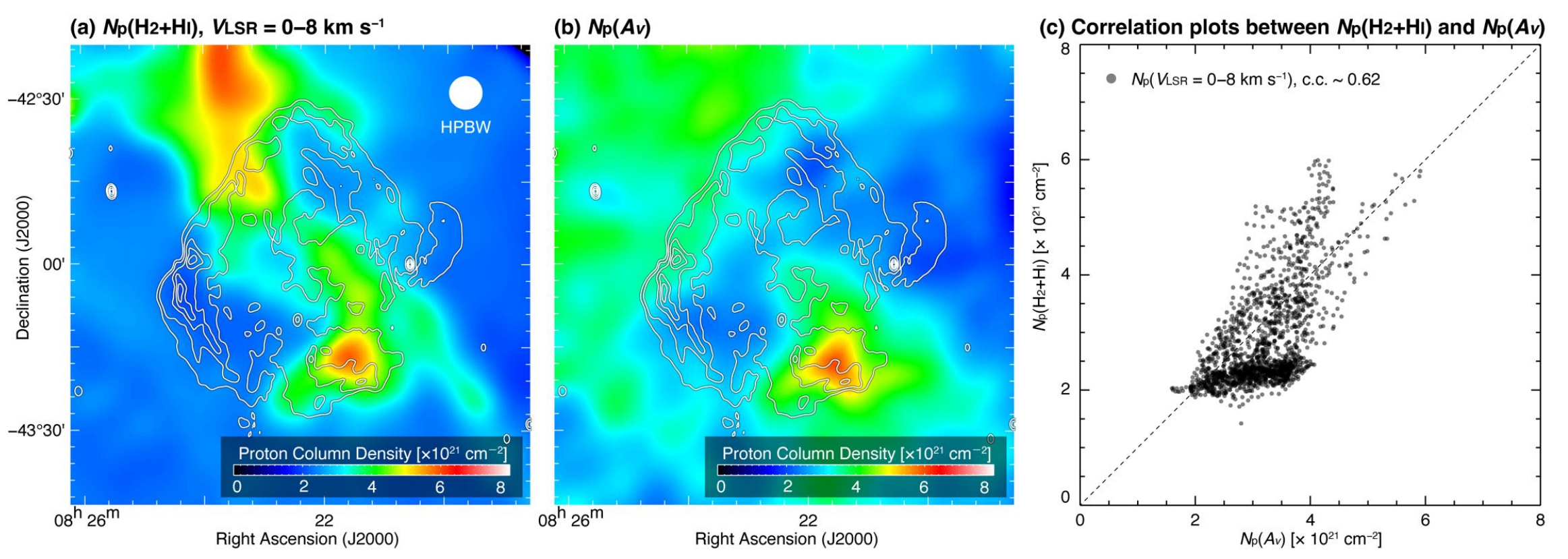}
\caption{(a) Column density maps of the ISM protons $N_{\rm p}$(H$_{2}$ $+$ H{\sc i}) estimated from $^{12}$CO($J$ = 1--0) and H{\sc i} integrated intensities. Velocity range of integrations are 0--8 km s$^{-1}$, (b) Column density map of $N_{\rm p}$ estimated from $A_{V}$\citep{2005PASJ...57..417D} (hereafter the "$N_{\rm p}$($A_{V}$)"). The superposed contours indicate the 1.4~GHz radio continuum as shown in Figure~\ref{fig2}, respectively. (c) Correlation plots between the $N_{\rm p}$($A_V$) and $N_{\rm p}$(H$_{2}$ $+$ H{\sc i}). The dashed line represents the equality of the two values. All the datasets used here are smoothed to a HPBW of $A_\mathrm{V}$ distribution ($\sim$6$'$) with a Gaussian function.}
\label{fig7}
\end{center}
\end{figure*}%

Figures \ref{fig6}(a) and \ref{fig6}(b) show the hardness ratio map of X-rays overlaid with $N_{\rm p}$(H$_{2}$ + H{\sc i}) contours of the {0--8}%3
~km~s$^{-1}$ and %16
{8--20}~km~s$^{-1}$ clouds. Figure \ref{fig6}c shows the correlation plots between the hardness ratio and $N_{\rm p}$(H$_{2}$ + H{\sc i}). The {0--8}%3
~km~s$^{-1}$ cloud shows spatial anticorrelation with $N_{\rm p}$(H$_{2}$ + H{\sc i}), indicating that the {0--8}%3
~km~s$^{-1}$ cloud is absorbing the soft X-rays. On the other hand, the %16
{8--20}~km~s$^{-1}$ cloud shows almost no correlation with the hardness ratio, indicating that the %16
{8--20}~km~s$^{-1}$ cloud is mostly not responsible for absorbing the soft X-rays. The correlation coefficient between the two quantities is { $\sim$0.66} %0.67 
in the {0--8}%3
~km~s$^{-1}$ cloud, and is  { $\sim$}0.0 in the %16
{8--20}~km~s$^{-1}$. 
%{We however cannot exclude the possibility of spatial correlation at ($\alpha_\mathrm{J2000}$, $\delta_\mathrm{J2000}) = (8^\mathrm{h}23^\mathrm{m}37\fs108$, $-42\arcdeg51\arcmin21\farcs45$), where the hardness ratio may become high due to some local mixing between the cloud and the SNR shell in the line of sight. *************************}
{We however cannot exclude the possibility of spatial correlation at ($\alpha_\mathrm{J2000}$, $\delta_\mathrm{J2000}) = (8^\mathrm{h}23^\mathrm{m}45^\mathrm{s}$, $-42\arcdeg48\arcmin$) as shown in dashed circle in Figure~\ref{fig6}(b), where the hardness ratio may become high due to some local mixing between the cloud and the SNR shell in the line of sight.}

Further, we calculate $N_{\mathrm p}(A_{V})$, the hydrogen column density from $A_{V}$ (Dobashi et al. 2005), by using the relationship {$N_{\mathrm p}(A_{V})\sim 1.87\times 10^{21} A_{V} (\rm cm^{-2})$} \citep{1978ApJ...224..132B}. %(Bohlin et al. 1978). 
Figure \ref{fig7} shows a scatter plot between  $N_{\mathrm p}(A_{V})$ and $N_{\rm p}$(H$_{2}$ + H{\sc i}) of the {0--8}%3
~km~s$^{-1}$ cloud. They are correlated with each other with a correlation coefficient of {$\sim$0.62}, %~0.6
supporting the trend in Figure \ref{fig6}. The anti-correlation in Figures \ref{fig6} and \ref{fig7} and lack of shock excitation in the {0--8}%3
~km~s$^{-1}$ cloud (see Figure \ref{fig3}a) are all consistent with that the {0--8}%3
~km~s$^{-1}$ cloud lies in front of the SNR and is not interacting with the SNR shock. This also explains the intensity decrease of the {H{\sc i}} emission toward the radio continuum shell as due to absorption in Figure~\ref{fig2}.

\subsubsection{Shock interaction and the ISM cavity}\label{discussion:shock interaction and the ISM cavity}
%\subsubsection{Shock interaction and expansion of the ISM shell}\label{discussion:shock interaction and expansion of the ISM shell}
We presented the signatures of the shock interaction in part of the CO gas in the %16
{8--20}~km~s$^{-1}$ cloud, which include the enhanced CO line intensity ratio $R_{\rm CO}$ and the broadened CO wing toward the northeast of the shell. $R_{\rm CO}$ is used for identifying the molecular gas associated with a SNR \citep{1998ApJ...505..286S}. %(Seta et al. 1998). 
The rotational state with $J$~=~2 has temperature 16~K from the ground level $J$~=~0, and a comparison of the state with $J$~=~1 state at 5~K provides an indicator of the rotational excitation. $R_{\rm CO}$ therefore gives a measure of the excitation state driven by temperature and/or density \citep[e.g.,][]{1998ApJ...505..286S, 2013ApJ...768..179Y}. %(e.g., Seta et al. 1998; Yoshiike et al. 2013). 
Toward Puppis A $R_{\rm CO}$ in the {0--8}%3
~km~s$^{-1}$ cloud is as low as ~0.5, while $R_{\rm CO}$ in the %16
{8--20}~km~s$^{-1}$ cloud toward the northeast of the SNR $R_{\rm CO}$ is as high as $\sim$0.8--1.1 (Figures \ref{fig3}a and \ref{fig3}b). This high ratio is likely due to the shock heating by the SNR because there is no other heating source. The CO wing is also a common signature of shock acceleration. In the middle{-}aged SNRs W28, W44 and IC 443, broad {$^{12}$CO} wings of 20--40~km~s$^{-1}$ are observed and are interpreted as due to the shock acceleration 
\citep[e.g.,][]{1979ApJ...232L.165D, 1977ApJ...216..440W, 1981ApJ...245..105W}. %(e.g., DeNoyer 1979; Wootten 1977, 1981). 
The present {$^{12}$CO} broad feature has a moderate velocity span of $\sim$5~km~s$^{-1}$ and the acceleration seems to be not so strong as in the other SNRs. Nonetheless, the feature is well recognized independently from the quiescent narrow component (Figure \ref{fig3}c) and is associated with the high $R_{\rm CO}$, lending support for the shock interpretation.

The ISM associated with the SNR shell is expanding by acceleration due to the supernova explosion and/or the stellar winds of the progenitor star. Puppis A is a remnant of a core-collapse explosion, and the progenitor was {likely} a high-mass star which produced strong stellar winds { \citep{1996ApJ...465L..43P, 1999ApJ...525..959Z, 2017MNRAS.464.3029R}.} 
%Petre et al. 1996, Zavlin et al. 1999, Reynoso et al. 2017
We showed that the CO gas in the %16
{8--20~km~s$^{-1}$ cloud show a cavity %arch-like distribution 
in p--v diagrams (Figures \ref{fig4}b and \ref{fig4}f).} %(Figures \ref{fig4}c and \ref{fig4}f), which is modelled by 10~km~s$^{-1}$ expansion. 
{Such %arch-like velocity 
cavity like distribution is a possible signature %of acceleration by 
due to explosive events \citep[e.g.,][]{1974ApJ...192..457C}.} %(e.g., Chevalier 1974). 
%We suggest that the expansion was driven by the stellar winds prior to the supernova over Myr timescale. 
The {0--8}%3
~km~s$^{-1}$ cloud also shows correlation with the radio continuum shell (Figure \ref{fig2}), but it is likely that the correlation is very close to the radio continuum emission, suggesting an absorption origin of the radio continuum emission that is in the background of the {0--8}%3
~km~s$^{-1}$ cloud.

A cloud associated with a SNR sometimes shows enhanced X-rays or radio continuum emission near and toward the SNR shell  \citep[e.g.,][]{2010ApJ...724...59S, 2013ApJ...778...59S, 2017JHEAp..15....1S}. %(e.g., Sano et al. 2010, 2013, 2017). 
We find such a trend in the %16
{8--20}~km~s$^{-1}$ CO cloud which shows peaks of the radio continuum emission and X-rays separated from the SNR by {a few tens of arcsec} %a few pc 
(Figure \ref{fig5}). This suggests that the %16
{8--20}~km~s$^{-1}$ cloud is heated up by the SNR. In addition, the H$\alpha$ emission shows anti-correlated distribution with the %16
{8--20}~km~s$^{-1}$ CO cloud (see also {Figure \ref{fig5}})%(see also Figure \ref{fig6})
, suggesting that the gas is ionized by the shock/winds and/or FUV photons product of X- ray ionization \citep[e.g.,][]{1983ApJ...267..603P}, see also the discussion in \citet{2008A&A...480..439P}. %(e.g., Prasad \& Tarafdar 1983, see also the discussion in Paron et al. 2008).
%参考文献の引き方確認, (e.g., Prasad & Tarafdar 1983, see also the 469 discussion in Paron et al. 2008).

\subsection{Distance, ISM Mass, and Age of the SNR Puppis~A}\label{discussion:distance and age}
An early VLA H{\sc i} study placed Puppis A at a distance of $\sim$ 2.2~kpc \citep{1995AJ....110..318R}. %(Reynoso et al. 1995). 
Additional hints supporting such distance were reported in a subsequent H{\sc i} survey towards the inner region of the SNR \citep{2003MNRAS.345..671R}. %(Reynoso et al. 2003). 
However, a more recent H{\sc i} absorption study with improved sensitivity and resolution (both spatial and spectral) covering the full SNR and its surroundings, in which single dish data were combined with interferometric data so as to ensure that large scale structures were sampled, revised the distance to be $\sim$ 1.3~kpc \citep{2017MNRAS.464.3029R}. %(Reynoso et al. 2017).
%Toward Puppis A, the H{\sc i} absorption was used to estimate the distance to be $\sim$2.2 kpc (Reynoso et al. 2003) and $\sim$1.3 kpc (Reynoso et al. 2017), while the CO emission was not used in the previous estimate. 
In the present work, we used the CO $J$ = 2--1 distribution obtained with NANTEN which covers the whole SNR at 1.3 arcmin resolution in conjunction with the H{\sc i} and X-rays, and identified the interaction signatures of the CO with the SNR shell. {The %16
{8--20}~km~s$^{-1}$ cloud has been confirmed to be interacting with the SNR.} {From Figure \ref{fig3}c, it is clear that the narrow component of the %16
%{8--20}~km~s$^{-1}$ cloud, which indicates the cloud systemic velocity, peaks at $\sim$10--11~km~s$^{-1}$. Following the Galactic rotation model by \citet{1989ApJ...342..272F}, this velocity corresponds to d~$\sim$$1.4~\pm~0.1$~kpc. Since we have demonstrated that Puppis~A is interacting with this cloud, hereafter we will adopt this distance for both the cloud and the SNR. This result is in excellent agreement with \citet{2017MNRAS.464.3029R} %Reynoso et al. (2017) 
%based on the H{\sc i} absorption.}
{8--20}~km~s$^{-1}$ cloud, which indicates the cloud systemic velocity, peaks at $\sim$10--11~km~s$^{-1}$.} {This result is in excellent agreement with \citet{2017MNRAS.464.3029R} %Reynoso et al. (2017) 
based on the H{\sc i} absorption.} {Following the Galactic rotation model by \citet{1989ApJ...342..272F}, this velocity corresponds to {$d$}~$\sim$$1.4~\pm~0.1$~kpc. Since we have demonstrated that Puppis~A is interacting with this cloud, hereafter we will adopt this distance for both the cloud and the SNR. }%This result is in excellent agreement with \citet{2017MNRAS.464.3029R} %Reynoso et al. (2017) based on the H{\sc i} absorption.}
The %16
{8--20}~km~s$^{-1}$ cloud has been confirmed to be interacting with the SNR and the distance {1.4~kpc} is supported.
%and the distance {1.4 $\pm$ 0.1} 
%2.2 kpc is supported. 
{Since the CO velocity range of the interacting cloud is narrower than the velocity range derived from the H{\sc i} absorption study by \citet{2017MNRAS.464.3029R}, %Reynoso et al. (2017), 
we are able to pinpoint the cloud systemic velocity with higher accuracy and thus improve the precision of the distance determination.}

We then derive interstellar proton density $n_{\rm p}$(H$_{2}$ + H{\sc i}) and the total mass of the interstellar protons by assuming a distance of {1.4 kpc.} In order to estimate the %mass of the CO clouds
{$\mathrm H_{2}$ molecular mass of the CO clouds} $M_\mathrm{CO}$, we used the following equation:
\begin{eqnarray}
M_\mathrm{CO} = \mu m_{\mathrm H} \Sigma_{i} [{D^{2}} \Omega N_{i}(\mathrm H_{2})],
\label{eq4}
\end{eqnarray}
where $\mu$ is the mean molecular weight, $m_{\mathrm H}$ the mass of the atomic hydrogen cloud, $D$ the distance to the SNR, $\Omega$ the solid angle of a square pixel, and $N_{i}(\mathrm H_{2})$ the {molecular} hydrogen column density of each pixel $i$. 
{We used $\mu$ = 2.8 including the contribution by Helium. }
%We used $\mu$ = 2.8 to account for a ratio of helium abundance to molecular hydrogen of 20\%. }
%We used $\mu$ = 2.8 to account for a helium abundance of 20\%. 
The hydrogen column density $N(\mathrm H_{2})$ is derived by using the equation ($X$ = 1.5 $\times$ 10$^{20}$ cm$^{-2}$ (K~km~s$^{-1}$)$^{-1}$, $N(\mathrm H_{2}) = X \cdot W(\mathrm{CO})$ (cm$^{-2}$)). We estimated the mass of molecular clouds within the radio shell extent (= shell radius + 1/2 shell thickness) to be $\sim${0.1} $\times$ 10$^{4}$ $M_{\odot}$. The mass of atomic hydrogen $M_{\mathrm{HI}}$ is derived by the equation~\ref{eq5}:%(5):
\begin{eqnarray}
M_{\mathrm{HI}} = m_{\mathrm H} {D^{2}} \Omega \Sigma_{i} [N_{i}(\mathrm{H}{\textsc{i}})],
\label{eq5}
\end{eqnarray}
where the $N(\mathrm{H}{\textsc{i}})$ is given by the optical-depth-corrected $N_{\mathrm p}(\mathrm{H}{\textsc{i}})$ (see {Section \ref{results:Hardness ratio Av map}}). %section 3.6). 
We thus estimated the mass of atomic clouds to be $\sim${1.3} $\times$ 10$^{4}$ $M_{\odot}$, which is about  {10} %20
times larger than that of the molecular clouds. The averaged number densities within the radio shell were estimated to $\sim${10}~cm$^{-3}$ for the molecular hydrogen $n(\mathrm{H_{2}})$, and $\sim${220}~cm$^{-3}$ for the atomic hydrogen $n(\mathrm{H}{\textsc{i}})$ by adopting the shell radius of $23\farcm8$ ($\sim${9.7}%9.67}
~pc) and the shell thickness of $6\farcm73$ ($\sim${2.7}%2.74}
~pc). Then, we derived the number density of the total interstellar protons $n$ to be 2 $\times n(\mathrm{H_{2}})$ + $n(\mathrm{H}{\textsc{i}})$ $\sim${230}~cm$^{-3}$.

The estimated age of Puppis~A is scattered in a large range of $\sim$2000--10000~yr in the literature \citep[e.g.,][]{1977ASSL...66...29C, 1979MNRAS.187..201C, 2015mbhe.confE..21A, 2020ApJ...899..138M}. %(e.g., Culhane et al. 1977; Caswell \& Lerche 1979; Aschenbach et al. 2015 ; Mayer et al. 2020). 
We briefly summarize the previous studies below.

\begin{figure*}[]
\begin{center}
\includegraphics[width=170mm,clip]{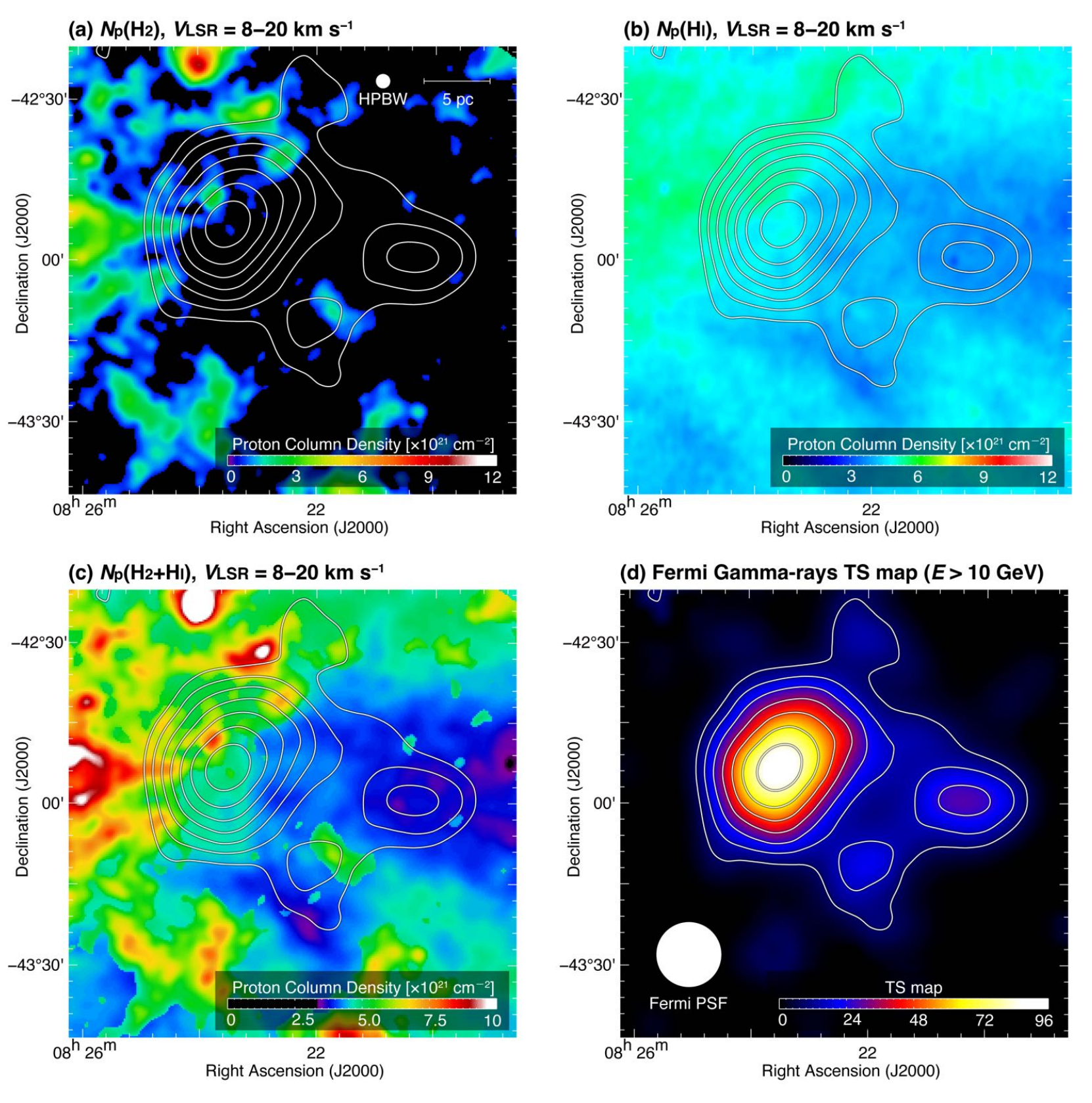}
\caption{(a--c) Distributions of the ISM proton column densities (a) $N_{\rm p}$(H$_{2}$), (b) $N_{\rm p}$(H{\sc i}), and (c) $N_{\rm p}$(H$_{2}$ + H{\sc i}). Velocity range of integrations are 8--20 km s$^{-1}$ for the each map. (d) Gamma-ray distribution  above 10GeV as shown in Figure \ref{fig1}.} 
%\citep{}. }
%(a\UTF{2013}c) Distributions of ISM proton column density, Np, are shown. All the data sets used for the three images are smoothed to have an HPBW of 8.′3, the same figure as the HPBW of H.E.S.S. (a) Np(H2) estimated from 12CO(J = 1\UTF{2013}0). (b) Np(Hi) estimated from Hi with correction of self-absorption. (c) Np(H2+Hi) estimated by summing up of Np (H2 ) and Np (H i). (d) TeV γ -ray distribution. Contours are plotted every 50 smoothed counts from 20 smoothed counts.
\label{fig8}
\end{center}
\end{figure*}%

%\clearpage%\clearpage
\begin{figure}[]
\begin{center}
\includegraphics[width=93mm]{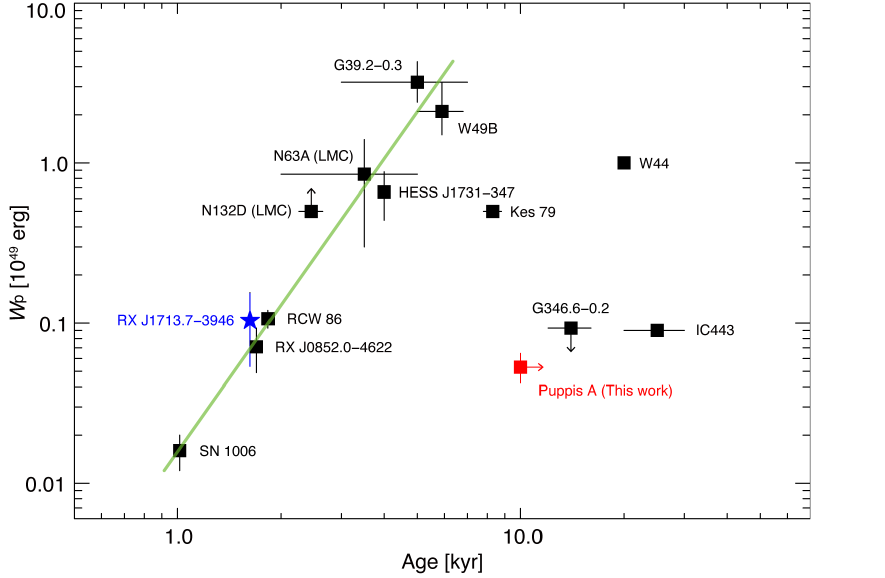}
\caption{Correlation plot between the age of SNRs and the total energy of {cosmic ray} protons $W_\mathrm{p}$ \citep{2021ApJ...919..123S, 2022ApJ...933..157S}.
%2022arXiv220513712S}. 
The green line indicates the linear regression of the double-logarithmic plot applying the least-squares fitting for data points with the ages of SNRs below 6 kyr. }
\label{fig9}
\end{center}
\end{figure}%

The ionization timescale is given as 4200~yr, 7000~yr, and 10000~yr at various places of the SNR \citep{2022A&A...661A..31M}. %(Mayer et al. 2022).
Other age estimates come from the proper motion of various parts of the SNR or the Sedov solution; 3700~$\pm$~300~yr \citep[the optical filament,][]{1988srim.conf...65W}%(the optical filament, \citet{1988srim.conf...65W})%Winkler et al. 1988)
, 4450~$\pm$~750~yr \citep[the CCO,][]{2012ApJ...755..141B}%(the CCO \citet{2012ApJ...755..141B})%, Becker et al. 2012)
, 1990~$\pm$~150~yr \citep[the ejecta with a decelerated model,][]{2015mbhe.confE..21A}%(the ejecta with a decelerated model, \citet{2015mbhe.confE..21A})%Aschenbach et al. 2015)
, 4600~yr \citep[the CCO,][]{2020ApJ...899..138M}%(the CCO, \citet{2020ApJ...899..138M})%Mayer et al. 2020)
%Aschenbach et al. 2015)%, 5000~yr (Sedov solution, Culhane et al. 1977)
, 4000--8000~yr \citep[Sedov solution,][]{1977ASSL...66...29C, 1979MNRAS.187..201C}. %(Sedov solution, \citet{1977ASSL...66...29C, 1979MNRAS.187..201C}).%Culhane et al. 1977, Caswell \& Lerche 1979).
%もう一度確認

Among all, we may adopt here the ionization age estimated from X-rays, which is defined as $\tau$ = $n_\mathrm{e} t_\mathrm{s}$ ($n_\mathrm{e}$ is the electron density and  $t_\mathrm{s}$ is the time since the emitting material was first struck by the shock wave \citep{2001AAS...19912619B, 2022A&A...661A..31M}. %Borkowski et al. 2001; Mayer et al. 2022). 
The age gives the upper limit to the timescale required to achieve the ionization state in the plasma, which in turn quantifies the degree of departure from collisional ionization equilibrium (CIE). It is correlated with the ISM identified in the present work (will be presented in a forthcoming paper), and we assume that the largest age 10000~yr is close to the true age of the SNR. Such a middle age seems to be consistent with the spectral properties of Puppis A, which has no synchrotron X-rays or bright TeV gamma{-}rays \citep[e.g.,][]{2015ARNPS..65..245F, 2021ApJ...908...22X}.%(e.g., Funk et al. 2015; Xiang \& Jiang 2021).

\subsection{Comparison with the gamma-ray distribution}\label{discussion:gammarays}
\subsubsection{Estimation of the cosmic ray energy from the hadronic gamma{-}rays}\label{discussion:CR energy}
In the previous works of the gamma-rays on SNRs, unfortunately not much attention has been paid on the details of the ISM distribution, and it has been often the case that the ISM density was too simply assumed to be uniform and as low as 1~cm$^{-3}$. Such assumptions obviously have no justification and lead to significant over estimation of the cosmic ray energy in order to compensate for lack of target protons in the p--p reaction. {In RX~J1713.7$-$3946, \citet{2021ApJ...915...84F} %Fukui et al. (2021) 
showed that the quantification of the hadronic gamma{-}rays is only possible by identifying the detailed ISM distribution.}
%It has been shown that the quantification of the hadronic gamma rays is possible only by identifying the detailed ISM distribution in RX~J1713.7$-$3946 by Fukui et al. (2021). This is easily understood by the linear dependence of the gamma ray flux on the ISM mass. Puppis~A is not an exception in this sense and very high $W_\mathrm{p}$ was estimated in the previous papers (e.g., Xin et al. 2017), whereas these approaches are not justified by the radio observations. The present study derived the ISM distribution in detail and allows us to investigate the issue quantitatively.

Figure \ref{fig8} shows that the GeV gamma{-}ray distribution has spatial correspondence with the %16
{8--20}~km~s$^{-1}$ cloud, whereas the correspondence is not conclusive given the limited angular resolution of the gamma-rays. Into more detail, we find in Figure \ref{fig1}(b) that the intense thermal X-rays, which reflect shock heating, are located near and in the eastern half of the primary gamma-ray peak, and in Figure \ref{fig8} that the eastern half of the primary gamma{-}ray peak is overlapped with the high column density ISM protons. Such correspondence may be similar to the gamma{-}ray-ISM correlation in the middle-aged SNR W44 \citep[see Figure \ref{fig7} in][]{2013ApJ...768..179Y}, %(see Figure \ref{fig7} in \citet{2013ApJ...768..179Y})%Yoshiike et al. 2013)
where the ISM {peaks} toward one side of the gamma-ray shell. This suggests it possible that at least part of the gamma{-}rays in the primary peak may be due to the hadronic origin. We also find that the secondary gamma{-}ray peak in the northwest is located toward X-ray features with little ISM, which may be a separate object at a different distance.

If we assume that all the gamma{-}rays in Puppis A is of the hadronic origin, the total energy of the cosmic rays are estimated to be as follows \citep[e.g.,][]{2006A&A...449..223A}%(e.g., Aharonian et al. 2006)
;
\begin{eqnarray}
%W_\mathrm{{p}}^{tot} \approx t_\mathrm{pp \rightarrow \pi_{0}} \times L_{\gamma},
W_\mathrm{{p}}^{tot} \sim t_\mathrm{pp \rightarrow \pi_{0}} \times L_{\gamma},
\label{eq6}
\end{eqnarray}
where $ t_\mathrm{pp \rightarrow \pi_{0}} {\sim}%{\approx} 
4.5 \times 10^{13} \left(\frac{n}{100~\mathrm{cm}^{-3}}\right)^{-1}$ s, is the characteristic cooling time of the protons and $ L_{\gamma}$ is the gamma-ray luminosity derived by \citet{2017ApJ...843...90X}. {An upper limit for $W_{\rm p}$ is given as follows:}
%(Xin et al. 2017). Assuming the distance is {1.4}~kpc, an upper limit for $W_\mathrm{p}$ is given as follows;
\begin{eqnarray}
% {W_\mathrm{p}~{\approx} %\lesssim 
{W_\mathrm{p}~{\sim} %\lesssim 
1.21 \times 10^{48} \left(\frac{n_{\mathrm p}}{100~\mathrm{cm}^{-3}}\right)^{-1} \left(\frac{d}{1.4~\mathrm{kpc}}\right)^{2} \mathrm{erg}.}%,
%W_\mathrm{p} \lesssim 2.2 \times 10^{13} \left(\frac{n}{100~\mathrm{cm}^{-3}}\right)^{-1} \left(\frac{d}{2.2~\mathrm{kpc}}\right)^{2} \mathrm{erg},
% \lessapprox
\label{eq7}
\end{eqnarray}
{If we adopt $n_\mathrm{p}$ =  {230~cm$^{-3}$} and $d$ =  {1.4~kpc}, we find $W_\mathrm{p}~{\sim}%{\approx} %\lesssim 
{5.3 \times 10^{47}}$ erg.} 
%, which corresponds to {$\sim$}
%$\approx$%$\lesssim$  
%{0.05\%}%0.2\% (2.2 kpc)
%of the total kinetic energy released by a SNe. 
Figure \ref{fig9} shows a plot between the SNR age and $W_\mathrm{p}$ \citep{2021ApJ...919..123S, 2021ApJ...923...15S, 2022ApJ...933..157S}.
%2022arXiv220513712S}%(Sano et al. 2021ab, 2022)
, and indicates that Puppis~A is located in a position consistent with the cosmic{-}ray escaping phase suggested by the other more than ten SNRs \citep{2022ApJ...933..157S}.%2022arXiv220513712S}.%(Sano et al. 2022).
% < ,  \lessapprox

\subsubsection{Escape of the CR protons}\label{discussion:escape}
We consider a significant part of the CR protons have escaped from the SNR shell in the time scale 10$^{4}$~yr. The cosmic-ray diffusion length $l_\mathrm{diff}$ can be described as \citet{2009MNRAS.396.1629G}%(Gabici et al. 2009)
\begin{eqnarray}
l_\mathrm{diff} = \sqrt{4 D(E) t_\mathrm{age}}.
% \lessapprox
\label{eq8}
\end{eqnarray}
Here $D(E)$ is {the} diffusion coefficient in units of cm$^{-2}$ s$^{-1}$, $t_\mathrm{age}$ is the age of SNR in units of second. The diffusion coefficient $D(E)$ can be written using the particle energy $E$ and the magnetic field strength $B$ as
\begin{eqnarray}
D(E) = 3 \times 10^{26} \left(\frac{E}{10~\mathrm{MeV}}\right)^{0.5} \left(\frac{B}{3~\mu\mathrm{G}}\right)^{-0.5}.
% \mathrm{cm^{-2} s^{-1}}
\label{eq9}
\end{eqnarray}
Here, by adopting $t_\mathrm{age} >$ 10000~yr, $E$ = 10~GeV, $B <$ 20~$\mu$G \citep{2018MNRAS.477.2087R}%(Reynoso et al. 2018)
, we obtain the cosmic-ray diffusion length $l_\mathrm{diff} >$ 22~pc. The length becomes large with $E$ and is larger than the SNR radius $\sim${10~pc}, supporting the escape of the cosmic rays.  {The} gamma{-}ray energy spectrum of Puppis~A which {peaks} at the GeV range is also consistent with the escape \citep[e.g.,][]{2015ARNPS..65..245F, 2021ApJ...908...22X}. %(e.g., Funk et al. 2015; Xiang \& Jiang 2021). 
\citet{2022MNRAS.510.2277A} %Araya et al. (2022) 
suggested that a nearby gamma{-}ray source 4FGL~J0822.8$-$4207 may be an object which is illuminated by the escaping cosmic rays from Puppis~A. An alternative idea is that the object is illuminated by cosmic rays produced in the star forming region. In future, we need a detailed comparison of the interstellar medium and gamma{-}ray distribution in order to explore the issue further.

\section{Conclusions}\label{conclusions}
We have investigated the ISM toward Puppis~A and identified that the clumpy CO clouds and the H{\sc i} gas in a velocity range of 8--20~km~s$^{-1}$ are associated with the SNR shell. Following this identification, we have examined the shock cloud interaction and the gamma{-}ray emission mechanism as well as the escaping cosmic rays from the SNR. The main conclusions are summarized below.
\begin{enumerate}
\item The CO and H{\sc i} distribution toward Puppis~A has two velocity ranges, i.e., 0--8~km~s$^{-1}$ %(3~km~s$^{-1}$ cloud) 
and {8--20~km~s$^{-1}$} %8--20~km~s$^{-1}$ (16~km~s$^{-1}$ cloud) 
as are inferred from their spatial distribution.
\item The %16
{8--20}~km~s$^{-1}$ cloud consists of small clumps which are distributed around the SNR shell, and shows enhanced $^{12}$CO $J$ = 2--1/1--0 ratio of 0.8--1.1, indicating its higher excited state toward the northeastern part of the SNR shell, where the %16
{8--20}~km~s$^{-1}$ cloud 
also shows a moderate wing-like feature of 5~km~s$^{-1}$ velocity span. The high ratio and broad feature suggest the gas is shock-excited/accelerated by the interaction with the SNR shell. Based on these properties we identify that the %16{11}~km~s$^{-1}$ 
{8--20}~km~s$^{-1}$ cloud  %cloud 
is interacting with the SNR. These signatures of the interaction support the distance {1.4~kpc}.
\item {The 0--8~km~s$^{-1}$~cloud} %The cloud at 3~km~s$^{-1}$} %The 3~km~s$^{-1}$ cloud 
is elongated from the southwest to the northeast across the SNR, and shows uniform and low $^{12}$CO $J$ = 2--1/1--0 ratio of 0.5 which is consistent with no extra excitation. {The 0--8~km~s$^{-1}$~cloud} %The cloud at 3~km~s$^{-1}$} %The 3~km~s$^{-1}$ cloud 
is associated with the lower hardness ratio of X{-}rays and shows intensity depression similar to the radio continuum shell. These properties suggest the cloud is located in front of the SNR, and the cloud is not physically interacting with the shell.
\item The H{\sc i} mass of the %16
{8--20}~km~s$^{-1}$ cloud was calculated to be $\sim$10$^{4}$~$M_{\odot}$, and the 
%CO cloud mass is about {1/10} of the H{\sc i} mass. 
 {$\rm H_2$ molecular mass of the CO cloud is about 1/10 of the atomic hydrogen mass.}
Based on the mass we estimate the cosmic{-}ray proton energy $W_\mathrm{p}$ to be $\sim$${5.3 \times 10^{47}}$~erg. 
%  to be $times 10^{48}$~erg. based ~ ->> Based
We also argue for an age of the SNR to be 10000~yr, the longest value derived from the X{-}ray ionization timescale. Then, we find that Puppis~A is placed in the cosmic{-}ray escaping phase in an age--$W_\mathrm{p}$ plot constructed by 13 SNRs.
\item Puppis~A shows shell like %X-ray 
{radio continuum} distribution similar to another middle-aged SNR W44. It seems %the 
{that} both SNRs {have} a relatively large ISM cavity of {10--15} %15--20} 
pc radius. 
{The primary gamma-ray peak is overlapped with the high column density ISM protons, suggesting that at least part of the gamma-rays in the primary peak %may be 
is due to the hadronic origin.}
The gamma{-}rays have their peak on one side of the shell, while the gamma{-}rays from the inside is not dominant compared to the shell. The angular resolution of the gamma{-}rays is however not high enough to explore further details of the gamma{-}ray properties including the contribution of the leptonic component, which must await future high-resolution observations with %CTA
{the Cherenkov Telescope Array}.
\end{enumerate}

\section*{Acknowledgements}
The authors acknowledge Dr. Yuliang Xin for providing gamma-ray maps of Puppis A appeared in \citet{2017ApJ...843...90X}. %Xin et al. (2017). 
The NANTEN project is based on a mutual agreement between Nagoya University and the Carnegie Institution of Washington (CIW). We greatly appreciate the hospitality of all the staff members of the Las Campanas Observatory of CIW. We are thankful to many Japanese public donors and companies who contributed to the realization of the project. The Australia Telescope Compact Array (ATCA) and the Parkes radio telescope are parts of the Australia Telescope National Facility which is funded by the Australian Government for operation as a National Facility managed by CSIRO. This study was based on observations obtained with XMM-Newton, an ESA science mission with instruments and contributions directly funded by ESA Member States and NASA. The scientific results reported in this article are based on data obtained from the Chandra Data Archive. This research has made use of the software provided by the Chandra X-ray Center (CXC) in the application packages CIAO (v 4.12). This work was supported by JSPS KAKENHI Grant Number JP21H01136 (H. Sano). {E.M.R. is member of the Carrera del Investigador Cient\'\i fico of CONICET, Argentina, and is partially funded by grant PIP 112-201701-00604CO.}
%\clearpage
\begin{figure}[]
\begin{center}
\includegraphics[width=80mm,clip]{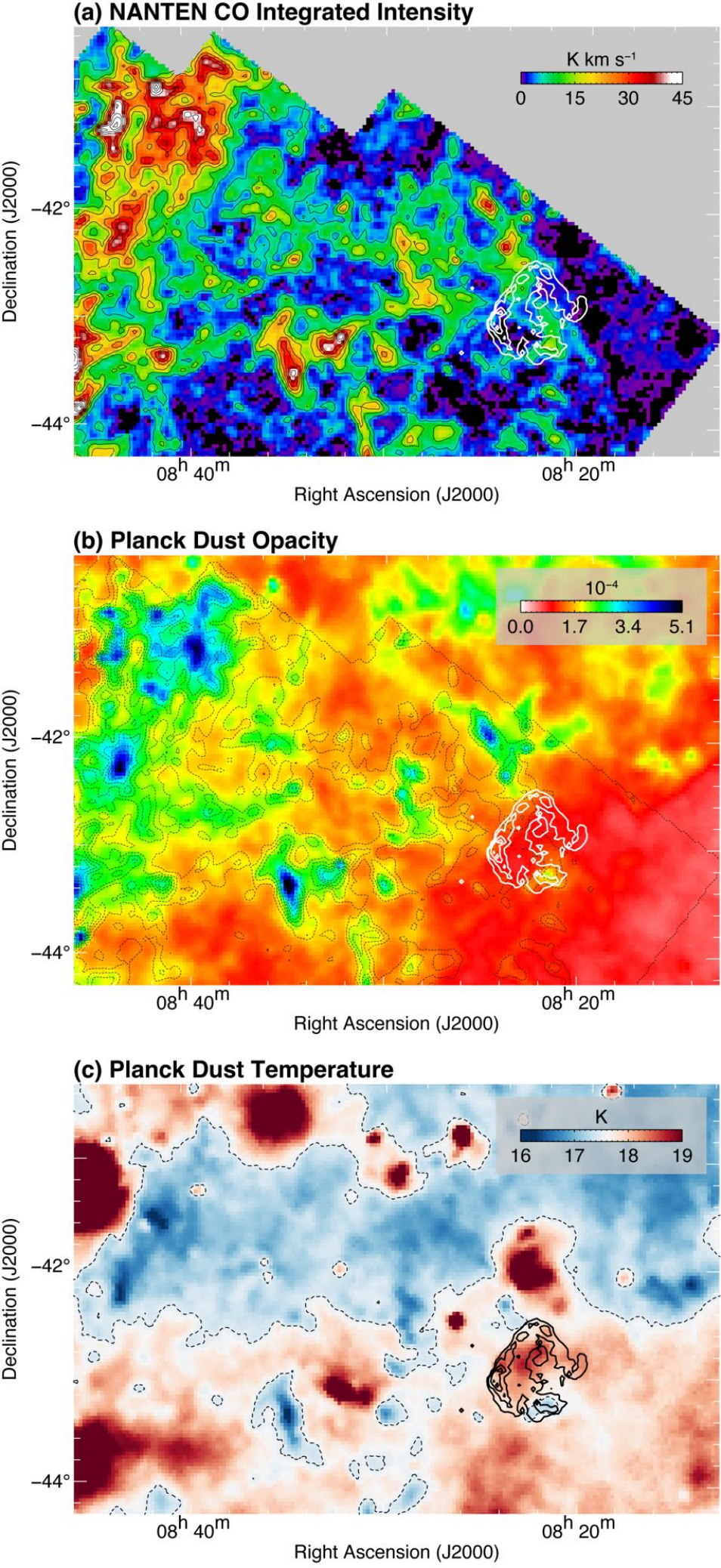}
\caption{Maps of (a) NANTEN $^{12}$CO($J$ = 1--0) integrated intensity \citep{2001PASJ...53.1025M}, (b) Planck dust opacity, and (c) Planck dust temperature \citep{2014A&A...571A..11P}. The integration velocity range of CO is from $-10$ to $80$ km s$^{-1}$. The superposed black contours in Figures \ref{figa1}(a) and \ref{figa1}(b) indicate the CO intensity, whose contour levels are from 6 K km s$^{-1}$ with the 6 K km s$^{-1}$ intervals. The dashed contours in Figure \ref{figa1}(c) represents the dust temperature of 17.5~K. The remaining superposed contours indicate the 1.4~GHz radio continuum and the contour levels are the same as shown in Figure \ref{fig2}.}
\label{figa1}
\end{center}
\end{figure}%

%\clearpage
\begin{figure}[]
\begin{center}
\includegraphics[width=80mm,clip]{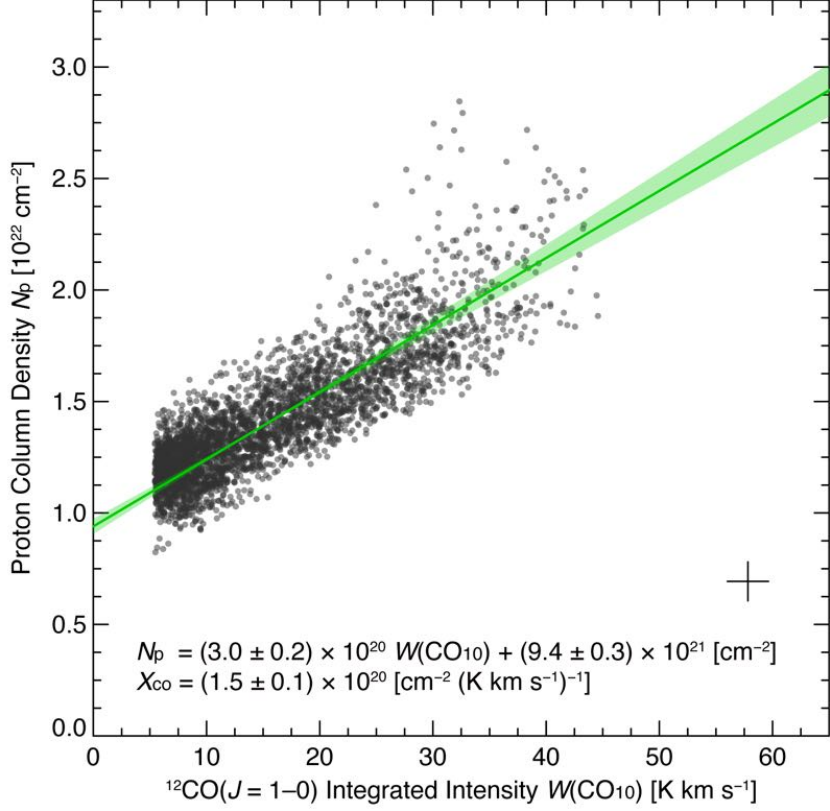}
\caption{Correlation plot between the $^{12}$CO($J$ = 1--0) integrated intensity $W$(CO$_{10}$) and the proton column density $N_\mathrm{p}$ derived using the Planck dust opacity and equation (\ref{eq10}). The typical error is shown in the right bottom corner. The linear regression by $\chi^2$ fitting is shown by the solid green line. The shaded area represents error for the fitting (see the text).}
\label{figa2}
\end{center}
\end{figure}%

{\software{IDL Astronomy User's Library \citep{1993ASPC...52..246L}, MIRIAD \citep[][]{1995ASPC...77..433S}, CIAO \citep[v 4.12:][]{2006SPIE.6270E..1VF}, CALDB \citep[v 4.9.1][]{2007ChNew..14...33G}, SAS \citep[v19.1.0:][]{2004ASPC..314..759G}, ESAS \citep{2008A&A...478..575K}, HEAsoft \citep[v6.28:][]{2014ascl.soft08004N}.}
%IDL Astronomy User’s Library (Landsman 1993), MIRIAD (Sault et al. 1995), CIAO (v 4.12: Fruscione et al. 2006), CALDB (v 4.9.1 Graessle et al. 2007), SAS (v19.1.0: Gabriel et al. 2004), ESAS (Kuntz & Snowden 2008), HEAsoft (v6.28: NASA High Energy Astrophysics Science Archive Research Center 2014). CIAO と SAS, CALDB, HEAsoft は、本文に出てくるバージョンを確認.
%***** ***** ***** ***** ***** ***** ***** ***** ***** ***** ***** ***** ***** ***** ***** ***** ***** ***** ***** ***** ***** ***** ***** ***** ***** ***** ***** ***** ***** ***** ***** ***** ***** 
\facilities{NANTEN, NANTEN2, Australia Telescope Compact Array (ATCA), Parkes, Chandra, XMM-Newton, and Fermi-LAT.}}

\section*{Appendix A: Determination of CO-to-H$_{2}$ Conversion Factor}
To derive the specific CO-to-H$_{2}$ conversion factor $X$ (also known as $X$-factor) in the Vela region, we used maps of the NANTEN $^{12}$CO($J$ = 1--0) integrated intensity $W$(CO), Planck dust opacity $\tau_{353}$ at the frequency of 353~GHz, and Planck dust temperature, following the method presented in \citet{2017ApJ...838..132O}.%Okamoto et al. (2017).

Figures \ref{figa1}(a), \ref{figa1}(b), and \ref{figa1}(c) show the large-scale (20 degree$^{2}$) maps of the NANTEN $^{12}$CO($J$ = 1--0) integrated intensity, Planck dust opacity $\tau_{353}$, and Planck dust temperature, respectively. We found clumpy and filamentary molecular clouds around Puppis A, which spatially correspond to the regions of large optical depth in $\tau_{353}$ (see Figure {\ref{figa1}b} image and black contours). 
%\ref{fig7}b  image and black contours). 
We also noted that the southern half of the map shows a high-dust temperature greater than 17.5~K, which was likely caused by shock-heating due to the overlapping Vela SNR \citep[see also][]{2001PASJ...53.1025M}. %(see also \citet{2001PASJ...53.1025M}).%Moriguchi et al. 2001).

According to \citet{2017ApJ...838..132O}%Okamoto et al. (2017)
, the total interstellar proton column density $N_\mathrm{p}$ is given by
\begin{eqnarray}
N_{\mathrm p} = 9.0 \times 10^{24}~{\tau_{353}}^{1/1.3}.
\label{eq10}
\end{eqnarray}
where the non-linear term of 1/1.3 is known as the dust-growth factor \citep[e.g.,][]{2013ApJ...763...55R, 2015ApJ...798....6F, 2017ApJ...838..132O}.%(e.g., Roy et al. 2013; Fukui et al. 2015; Okamoto et al. 2017).

Figure \ref{figa2} shows a scatter plot between $W$(CO) and {$N_{\mathrm p}$} derived using the equation (\ref{eq10}). We fitted the data points of CO with 5$\sigma$ or higher significance using the MPFITEXY procedure which provides the slope, intercept, and reduced-$\chi^2$ values of the linear function {\citep{2010MNRAS.409.1330W}}. %{(Williams et al. 2010).} 
We found the best fit values with reduced-$\chi^2$ $\sim$1.01 (degree of freedom = 3500) when we used only the data points with the dust temperature below 17.5~K. We finally obtained the slope of (3.0~$\pm$~0.2) $\times 10^{20}$ cm$^{-2}$ (K~km~s$^{-1}$)$^{-1}$ and the intercept of (9.4~$\pm$~0.3) $\times 10^{21}$ cm$^{-2}$. From equations (\ref{eq1}) and (\ref{eq2}), we then obtained the CO-to-H$_{2}$ conversion factor $X$ = 1.5 $\times$ 10$^{20}$ cm$^{-2}$
(K~km~s$^{-1}$)$^{-1}$ for the Vela region.
%\chi , Chi

\section*{Appendix B: Radial Profile for the Radio Continuum Shell}
\begin{figure}[t]
\begin{center}
:\includegraphics[width=\linewidth,clip]{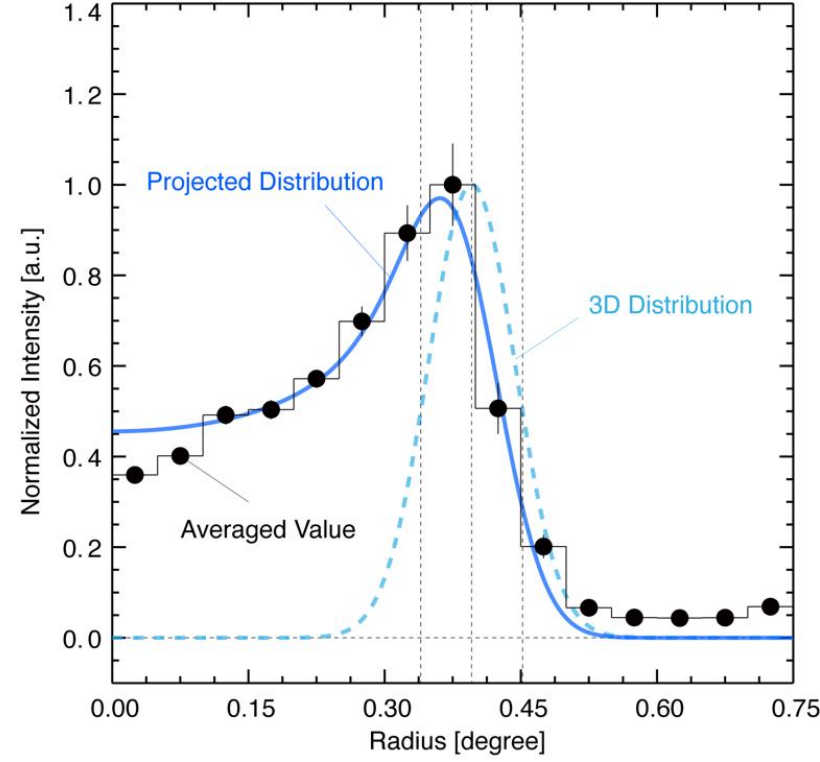}
\caption{Radial profiles of the radio continuum, centered at ($\alpha$, $\delta$) $=$ (125\fdg59, $-$43\fdg03). The black steps with filled circles indicate averaged values of radio continuum at each annulus. The dashed line represents the three-dimensional Gaussian distribution and the solid line represents its projected distribution derived by the least-squares fitting (see the text). The vertical dashed lines indicate best-fit values of radius and ranges of shell thickness. }
\label{figa3}
\end{center}
\end{figure}

To derive the apparent diameter and shell thickness of the radio continuum shell, we fitted its radial profile using a three-dimensional spherical shell with a Gaussian function $F(r)$,
\begin{eqnarray}
F(r) = A \exp[-(r-r_{0})^2 / 2\sigma^2],
\label{eq11}
\end{eqnarray}
where $A$ is a normalization of the Gaussian function, $r_{0}$ is the shell radius in units of arcmin, and $\sigma$ is the standard deviation of the Gaussian function in units of arcmin. We first fitted the radial profile by moving the original position around the geometric center of the SNR, and we obtained the central position of the shell as ($\alpha_\mathrm{J2000}$, $\delta_\mathrm{J2000}$) = $(125\fdg59, -43\fdg03)$ with the minimum chi-square value of the least-squares fitting. {Figure \ref{figa3}} shows the radial profile of the radio continuum, centered at ($\alpha_\mathrm{J2000}$, $\delta_\mathrm{J2000}$) = $(125\fdg59, -43\fdg03)$. We derived the shell radius $r_{0}$ of $0\fdg40\pm0\fdg08$ and thickness of $0\fdg11\pm0\fdg19$ as the best-fit parameters, where the shell thickness is defined as the FWHM of the Gaussian function or $2\sigma \sqrt{2 \ln(2)}$.

%%%%%%%%%%%%%%%%%%%%%%%%%%%%%%%%%%%%%%%%%%%%%%%%%%%%%%%%%%%%%%%%%%%%%%%%
%%%%%%%%%%%%%%%%%%%%%%%%%%%%%%%%%%%%%%%%%%%%%%%%%%%%%%%%%%%%%%%%%%%%%%%%
%%%%%%%%%%%%%%%%%%%%%%%%%%%%%%%%%%%%%%%%%%%%%%%%%%%%%%%%%%%%%%%%%%%%%%%%

%\begin{figure*}[]
%\begin{center}
%:\includegraphics[width=\linewidth]{_appendix_RCimgRCcont_radialprofile_RCgamma0p501_SB32p2kpc.pdf}
%%\includegraphics[width=150mm,clip]{appendix_RCimgRCcont_radialprofile_RCgamma0p501_SB32p2kpc.pdf}
%\caption{Left panel : Map of 1.4 GHz radio continuum from the ATCA $\&$ Parkes \citep{2017MNRAS.464.3029R}. The contour levels are the same as shown in Figure \ref{fig2}. Right panel: Radial profiles of the radio continuum, centered at ($\alpha$, $\delta$) $=$ (125\fdg59, $-$43\fdg03). The black steps with filled circles indicate averaged values of radio continuum at each annulus. The dashed line represents the three-dimensional Gaussian distribution and the solid line represents its projected distribution derived by the least-squares fitting (see the text). The vertical dashed lines indicate best-fit values of radius and ranges of shell thickness. }
%\label{figa1}
%\end{center}
%\end{figure*}%

\clearpage
%\newpage

\end{document}